# Generalist versus Specialist Vision Foundation Models for Ocular Disease and Oculomics

**Authors:** Yukun Zhou[1,2,3], Paul Nderitu[1,3], Jocelyn Hui Lin Goh[4,5,6], Justin Engelmann[1,3], Siegfried K. Wagner[1,3], Anran Ran[7], Hongyang Jiang[7], Lie Ju[1,3], Ke Zou[4,5], Sahana Srinivasan[4,5], Hyunmin Kim[1,3], Takahiro Ninomiya[1,3], Zheyuan Wang[8,9], Gabriel Dawei Yang[6,7], Eden Ruffell[1,3], Dominic Williamson[1,3], Rui Santos[10,11], Gabor Mark Somfai[10,11], Carol Y. Cheung[7], Tien Yin Wong[6,12,13], Daniel C. Alexander[2,14], Yih Chung Tham[4,5,6], Pearse A. Keane[1,3]

**Affiliations:**
1. Institute of Ophthalmology, University College London, London, UK
2. Hawkes Institute, University College London, London, UK
3. NIHR Biomedical Research Centre at Moorfields Eye Hospital NHS Foundation Trust, London, UK
4. Department of Ophthalmology, Yong Loo Lin School of Medicine, National University of Singapore, Singapore
5. Centre for Innovation and Precision Eye Health, Yong Loo Lin School of Medicine, National University of Singapore, Singapore
6. Singapore Eye Research Institute, Singapore National Eye Centre, Singapore.
7. Department of Ophthalmology and Visual Sciences, Chinese University of Hong Kong, Hong Kong, China.
8. Department of Computer Science and Engineering, Shanghai Jiao Tong University, Shanghai, China
9. MOE Key Laboratory of AI, School of Electronic, Information, and Electrical Engineering, Shanghai Jiao Tong University, Shanghai, China
10. Department of Ophthalmology, Stadtspital Zürich, Zurich, Switzerland
11. Spross Research Institute, Zurich, Switzerland
12. School of Clinical Medicine, Tsinghua Medicine, Tsinghua University, Beijing, China
13. Beijing Visual Science and Translational Eye Research Institute, Beijing Tsinghua Changgung Hospital, Beijing, China
14. Department of Computer Science, University College London, London, UK

**Corresponding author:**
Yukun Zhou, UCL Institute of Ophthalmology and UCL Hawkes Institute, United Kingdom. NIHR Biomedical Research Centre for Ophthalmology, Moorfields Eye Hospital NHS Foundation Trust. Email: yukun.zhou.19@ucl.ac.uk

**Abstract**

Medical foundation models, pre-trained with large-scale clinical data, demonstrate strong performance in diverse clinically relevant applications. RETFound, trained on nearly one million retinal images, exemplifies this approach in applications with retinal images. However, the emergence of increasingly powerful and multifold larger generalist foundation models such as DINOv2 and DINOv3 raises the question of whether domain-specific pre-training remains essential, and if so, what gap persists. To investigate this, we systematically evaluated the adaptability of DINOv2 and DINOv3 in retinal image applications, compared to two specialist RETFound models, RETFound-MAE and RETFound-DINOv2. We assessed performance on ocular disease detection and systemic disease prediction using two adaptation strategies: fine-tuning and linear probing. Data efficiency and adaptation efficiency were further analysed to characterise trade-offs between predictive performance and computational cost. Our results show that although scaling generalist models yields strong adaptability across diverse tasks, RETFound-DINOv2 consistently outperforms these generalist foundation models in ocular-disease detection and oculomics tasks, demonstrating stronger generalisability and data efficiency. These findings suggest that specialist retinal foundation models remain the most effective choice for clinical applications, while the narrowing gap with generalist foundation models suggests that continued data and model scaling can deliver domain-relevant gains and position them as strong foundations for future medical foundation models.

**Introduction**

Foundation models are large artificial intelligence models trained on broad and diverse datasets. With self-supervised techniques, foundation models can capture general-purpose data patterns that can be applied to diverse downstream applications [1,2]. By capturing medical domain features such as anatomical structures and pathological patterns, medical foundation models can be adapted to a wide range of clinically relevant applications [3–6]. These include disease detection, prognosis, and risk stratification across multiple specialities such as ophthalmology [7–9], radiology [10–12], and pathology [13–15].

The progress of medical foundation models has largely been enabled by methodological advances first pioneered in generalist vision foundation models. Self-supervised learning techniques such as contrastive learning (e.g. MoCo [16]), self-distillation (DINO [17], DINOv2 [18]), and masked image modelling (e.g. MAE [19]), as well as scalable model architectures (e.g. Vision Transformer [20]), have provided the building blocks for adapting foundation models to medical imaging [5,6,21–23]. Many medical foundation models are built directly on top of generalist models, with successive domain-specific pre-training used to capture medical features more effectively. A representative example is RETFound [7], a retinal foundation model pre-trained with MAE on approximately one million retinal images, which has substantially improved performance across both ocular and systemic prediction tasks. More recently, RETFound incorporated DINOv2 [18], a generalist foundation model developed with ~142 million images, into its pre-training pipeline, developing RETFound-DINOv2 with retinal images and further improving its effectiveness in ocular disease detection [24].

Parallel to these domain-specific efforts, generalist vision foundation models have continued to advance. The recent introduction of DINOv3 [25], trained on ~1.7 billion curated natural images with advanced strategies such as Gram Anchoring and high-resolution adaptation, represents a significant enhancement in open-source vision foundation models. Its scalability and robust representations raise an important question: to what extent can such generalist models be adapted to clinically relevant applications, and how do they compare with specialised medical foundation models such as RETFound? Similar questions have recently been explored [26], while addressing this in our context, covering more recent and broader foundation models (e.g. DINOv3, RETFound-DINOv2) for wide disease detection, will further clarify whether generalist foundation models may achieve clinically meaningful performance through fine-tuning or linear probing, in the absence of domain-specific pre-training.

In this work, we examined the adaptability of DINOv2 and DINOv3 to clinically relevant applications in retinal images and benchmarked them against the retina-specific foundation model, RETFound. Our evaluation covers two major categories of clinically relevant applications: ocular disease detection, using publicly available datasets for diabetic retinopathy, glaucoma, and other retinal disease detection, and systemic disease prediction from retinal images (termed as Oculomics [27,28]), where models are adapted to predict incident myocardial infarction, stroke, and heart failure within a three-year window. Results indicate that scaling generalist models such as DINOv2 and DINOv3 shows strong adaptability across diverse downstream tasks, while specialist RETFound models consistently demonstrate clear advantages. Besides, our analyses of data efficiency and adaptation efficiency highlight the trade-offs between performance and computational cost. Overall, these findings suggest that retinal foundation models remain the most effective choice for clinical applications with retinal images. Meanwhile, the competitive performance of newly proposed generalist models (e.g. DINOv3) demonstrates their strong potential as a robust basis for developing more efficient, clinically oriented foundation models.

**Method**

**Model description:** We included three series of vision foundation models: DINOv2, DINOv3, and RETFound, all using the Vision Transformer (ViT) architecture [20]. The DINOv2 series includes ViT-small, ViT-base, ViT-large, and ViT-giant, trained on LVD-142M, a dataset of 142 million natural images. The DINOv3 series comprises ViT-small, ViT-small+, ViT-base, ViT-large, ViT-huge+, and ViT-7b, trained on LVD-1689M, which contains 1.689 billion natural images. We did not include ViT-7B in this study as it requires significantly higher computational and data requirements for both fine-tuning and linear probing. The RETFound series includes RETFound-DINOv2 [24] and RETFound-MAE [7,24]: RETFound-DINOv2 was developed using 904K retinal images from the AlzEye cohort [29] initialised from DINOv2-ViT-large, while RETFound-MAE was built on MAE-ViT-large using the same 904K retinal images. More details regarding the model parameters and the release dates were shown in Figure 1.

**Datasets:** We evaluated model performance across two categories of clinically relevant applications: ocular disease detection and systemic disease prediction. For ocular disease detection, we included Kaggle APTOS-2019 (India), IDRiD (India) [30], and MESSIDOR2 (France) [31,32] for diabetic retinopathy detection, Papila (Spain) [33] and Glaucoma Fundus (South Korea) [34] for glaucoma detection, and Kaggle Retina for multiple retinal disease detection. Dataset characteristics are summarised in Supplementary Table 1. For systemic disease prediction, we followed the same data pipeline reported in the original RETFound study [7], which assessed 3-year prediction of cardiovascular diseases (including myocardial infarction, heart failure, and stroke) and neurodegenerative disease (including Parkinson's disease). The AlzEye cohort [29] was used for model adaptation and internal evaluations, while the UK Biobank data [35] was used for external evaluation. For each patient, we include a single left-eye image per visit to avoid visit-related bias. Detailed data characteristics are shown in Supplementary Table 2.

**Model adaptation to downstream tasks:** For disease detection and prediction, the ViT encoder was used to extract high-level retinal features, which were passed to a fully connected layer that produced probability distributions over disease categories. The number of output neurons matched the number of categories, and the category with the highest probability was taken as the predicted label. Two adaptation strategies were employed: fine-tuning and linear probing. Fine-tuning updated both the encoder and the fully connected layer, whereas linear probing trained only the fully connected layer. All images are resized to 256 × 256 with cubic interpolation. We follow the same data augmentation reported in RETFound [7] for model training, including random crop (lower bounds 20% of the whole image and upper bounds 100%) and resizing the cropped patches to 224 × 224, random horizontal flipping and image normalisation. AutoMorph [36] was used to pre-process the colour fundus images. Training was performed with a batch size of 24 for 50 epochs. The learning rate was warmed up linearly from 0 to $5 \times 10^{-4}$ over the first 10 epochs, followed by a cosine annealing schedule that decayed the rate from $5 \times 10^{-4}$ to $1 \times 10^{-6}$ over the remaining 40 epochs. After each epoch, performance was evaluated on the validation set, and the checkpoint with the highest sum of AUROC and F1 score was selected for internal and external evaluations.

**Model evaluation:** We conducted internal and external evaluations for the developed models. For internal evaluation, patient groups were split into training, validation, and test sets in a 55:15:30 ratio. The training set was used to optimise model parameters, the validation set to monitor convergence and select checkpoints, and the test set to assess final performance. For external validation, all available data were used to evaluate the saved checkpoint. All task performances were evaluated using three classification metrics: area under the receiver operating characteristic curve (AUROC), area under the precision-recall curve (AUPRC), and F1 score. For multiclass tasks, such as five-stage diabetic retinopathy grading, AUROC, AUPRC, and F1 scores were calculated for each class and then averaged to obtain overall performance. To evaluate the overall performance, we measured the model performance and ranks over the ten downstream tasks. We assigned performance ranks for each task and averaged them, with 1 indicating the best and 5 the worst. For model adaptation efficiency, we observed and compared the GPU memory required for both fine-tuning and linear probing for different models.

**Statistical analysis:** For each task, we trained a model and performed bootstrapping (n=200, repetitive sampling in full test size). The mean value and standard deviation across the 200 bootstraps were reported. The standard error was estimated as (standard deviation / $\sqrt{200}$), and 95% confidence intervals (CI) were obtained as 1.96 × standard error. We compute a cross-dataset p-value using a two-sided Wilcoxon signed-rank test to assess whether one model has a dataset-robust, consistently positive advantage over the other.

## Results

### Study framework

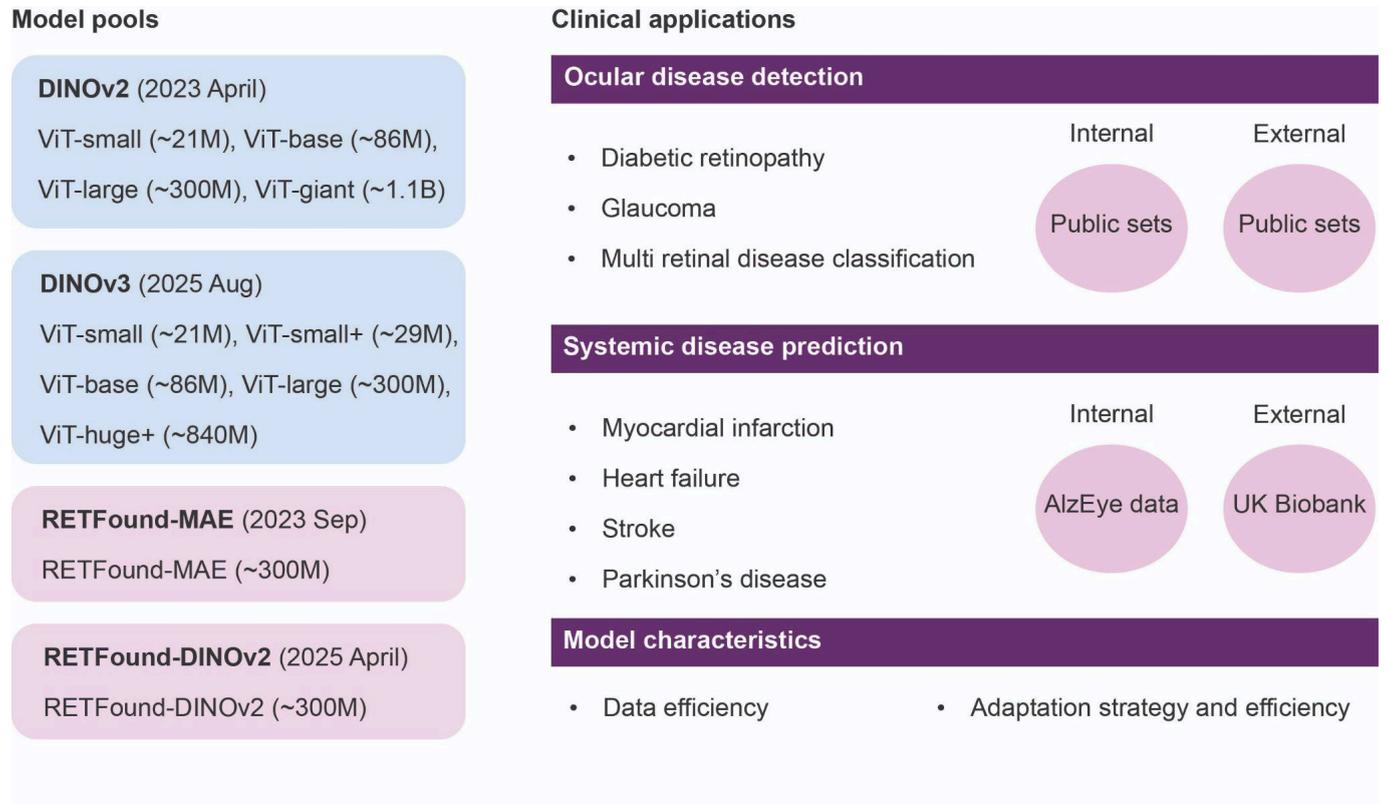

Figure 1. Evaluation and comparison framework for generalist foundation models DINOv2 and DINOv3, and retinal foundation model RETFound. Clinical applications include ocular disease detection and systemic disease prediction, organised using AlzEye data, UK Biobank, and publicly available datasets.

Figure 1 outlines the framework for evaluating and comparing DINOv2, DINOv3, and RETFound. DINOv2 and RETFound-MAE were developed in 2023, while DINOv3 and RETFound-DINOv2 were recently proposed. In this study, we evaluated models on clinical applications, including ocular disease detection and systemic disease prediction, and assessed their key characteristics, such as data efficiency and adaptation efficiency.

**DINOv2-ViT-giant achieved the best performance among the DINOv2 series**

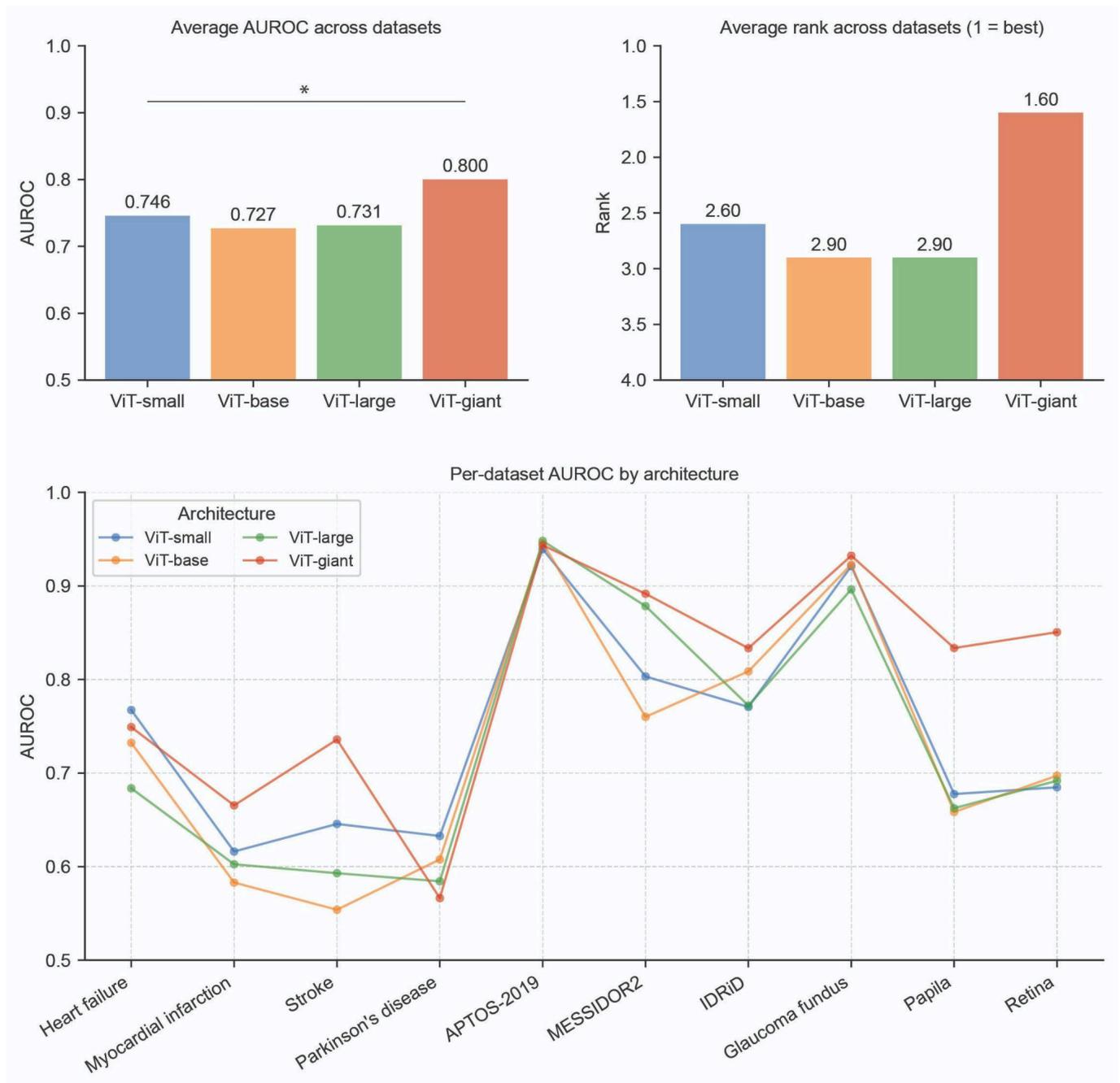

Figure 2. Fine-tuning performance comparison of four DINOv2 models to downstream tasks. The top subgraphs show the average performance and ranks across the ten tasks. DINOv2-ViT-giant significantly outperforms DINOv2-ViT-small according to the two-sided Wilcoxon signed-rank test ($p<0.05$). The X axis denotes the tasks, and the evaluation metric shows the AUROC value in the bottom subgraph. The 95% confidence interval is reported in the Supplementary Table 3.

As shown in Figure 2, DINOv2-ViT-giant consistently outperformed the others on multiple tasks, including stroke prediction and diabetic retinopathy detection on the MESSIDOR2 dataset. Its superiority was particularly evident on the Papila and Retina datasets. Across ten tasks, the average AUROC of ViT-small, ViT-base, ViT-large, and ViT-giant were 0.746, 0.727, 0.731, and 0.800, respectively (ViT-giant significantly outperformed ViT-small, the second-best performing model). The average rankings were

2.6, 2.9, 2.9, and 1.6, further highlighting the superior performance of DINOv2-ViT-giant. ViT-giant also achieved the best AUPRC and F1 score (Supplementary Figure 1). A similar trend was also observed under linear probe adaptation (Supplementary Figures 2 and 3). Full results are provided in Supplementary Table 3.

**DINOv3-ViT-large achieved the best performance among the DINOv3 series**

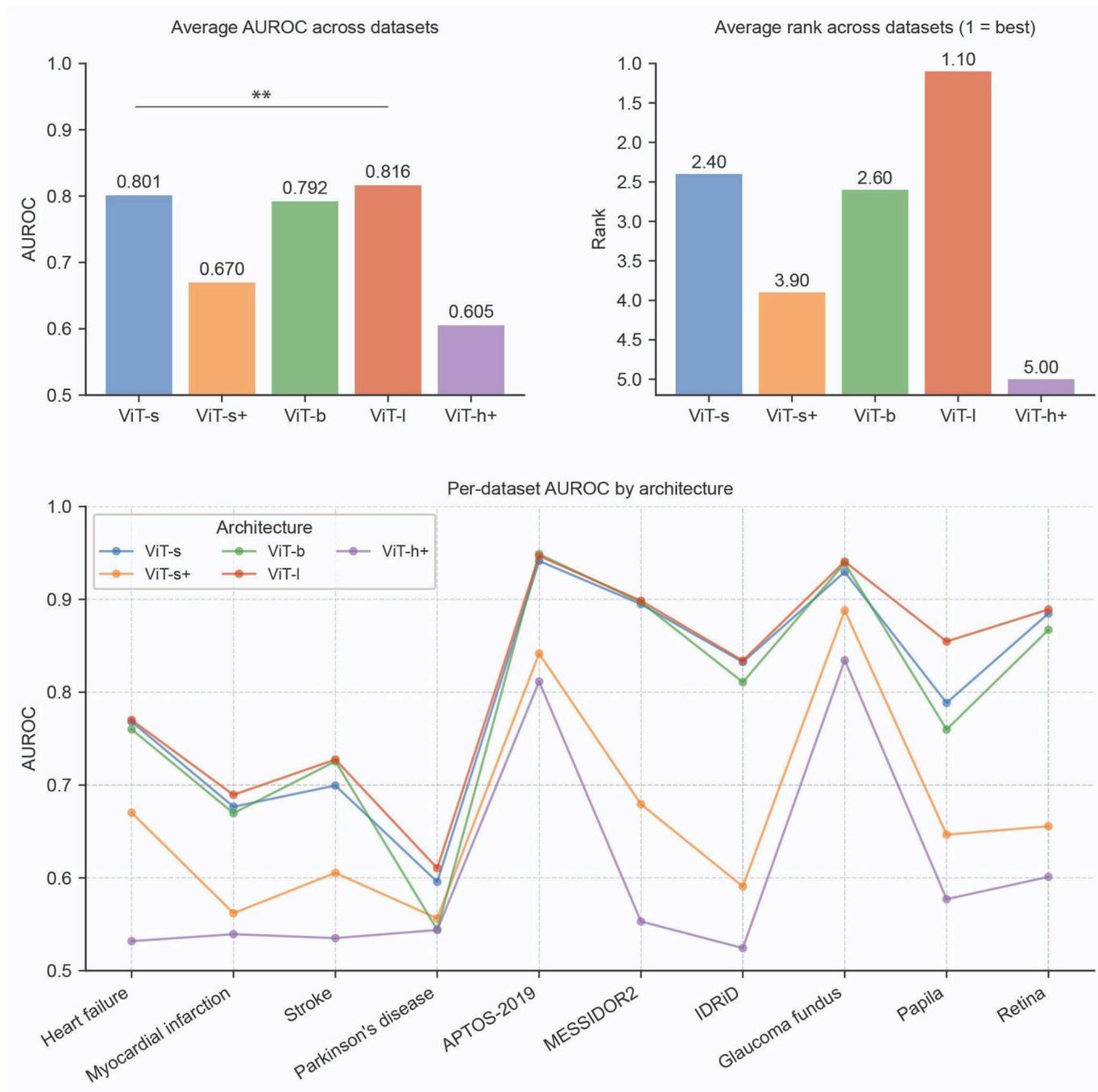

Figure 3. Fine-tuning performance comparison of four DINOv3 models to downstream tasks. The top subgraphs show the average performance and ranks across the ten tasks. DINOv3-ViT-large significantly outperforms DINOv3-ViT-small according to the two-sided Wilcoxon signed-rank test ($p<0.01$). The X axis denotes the tasks, and the evaluation metric shows the AUROC value in the bottom subgraph. ViT-s, ViT-b, and ViT-l denote ViT-small, ViT-base, and ViT-large, respectively. The 95% confidence interval is reported in the Supplementary Table 3.

As shown in Figure 3, ViT-small+ and ViT-huge+ underperformed compared to the other three models, with a substantial performance gap. ViT-small, ViT-base, and ViT-large demonstrated comparable results across most tasks, though ViT-large outperformed the others on datasets such as MESSIDOR2 and Papila. The average AUROC of the five DINOv3 models across ten tasks were 0.801, 0.670, 0.792, 0.816, and 0.605 respectively (ViT-large significantly outperformed ViT-small, the second-best performing model). The average rankings were 2.4, 3.9, 2.6, 1.1, and 5.0, respectively, confirming the superior performance of DINOv3-ViT-large. DINOv3-ViT-large also achieved the best AUPRC and F1 score (Supplementary Figure 4). A consistent trend was also observed under linear probe adaptation (Supplementary Figures 5 and 6). Full results are provided in Supplementary Table 3.

# RETFound-DINOv2 outperformed DINOv2 and DINOv3 in certain applications

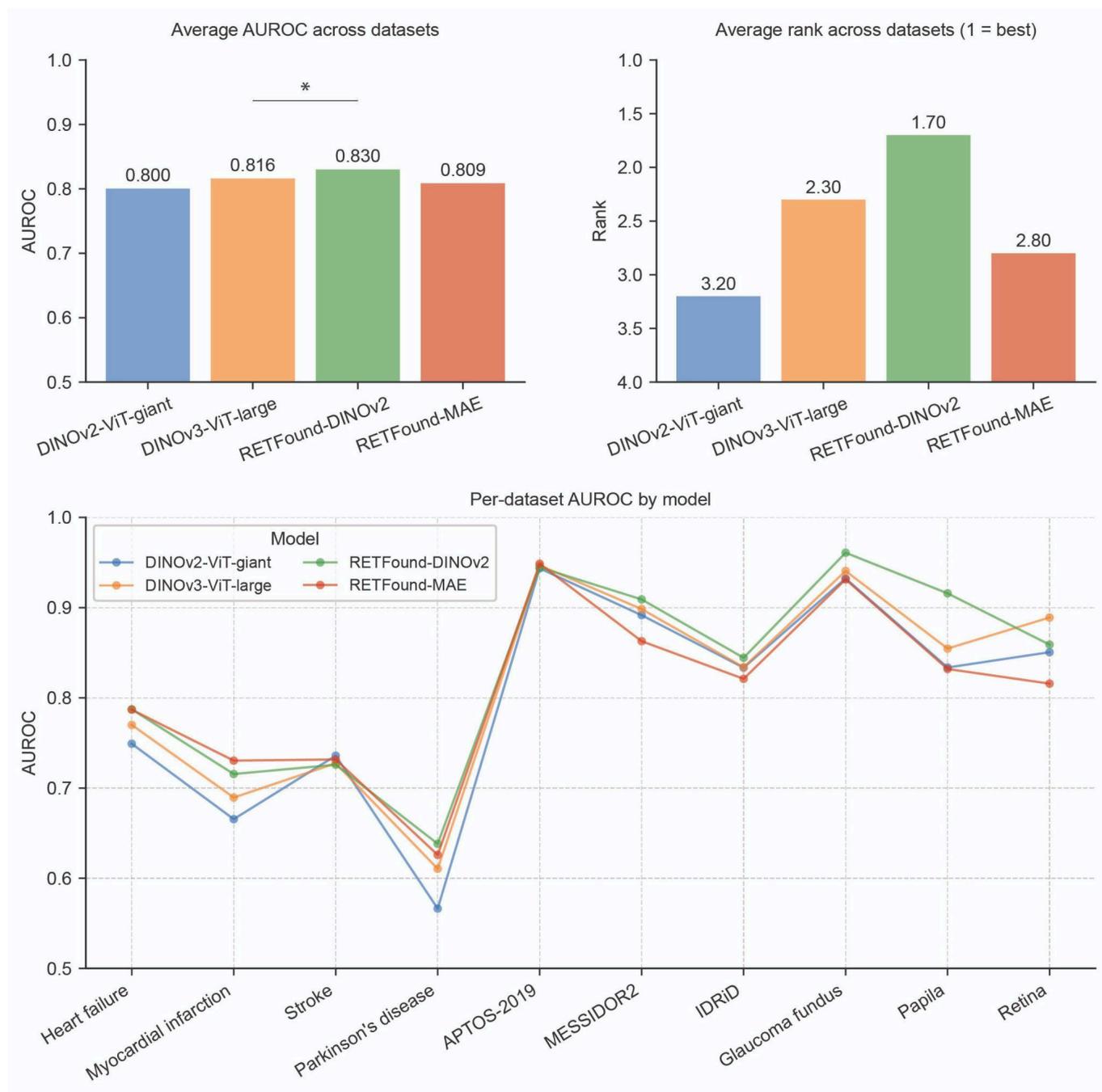

Figure 4. Fine-tuning performance comparison between DINOv2-ViT-giant, DINOv3-ViT-large, RETFound-DINOv2, and RETFound-MAE. The top subgraphs show the average performance and ranks across the ten tasks. RETFound-DINOv2 significantly outperforms DINOv3-ViT-large according to the two-sided Wilcoxon signed-rank test (p<0.05). The X axis denotes the tasks, and the evaluation metric shows the AUROC value in the bottom subgraph. DINOv2-ViT-giant (~1.1B parameters) is roughly 3.5 times larger than DINOv3-ViT-large, RETFound-DINOv2, and RETFound-MAE, each with ~300M parameters. The 95% confidence interval is reported in the Supplementary Table 3.

We compared the strongest baselines, DINOv2-ViT-giant and DINOv3-ViT-large, with RETFound models, which were pre-trained on retinal images. As shown in Figure 4, all four models performed comparably on most tasks, while RETFound-DINOv2 achieved higher AUPRC and F1 score on IDRiD, Glaucoma Fundus, and Papila datasets (Supplementary Figure 7). The average AUROC performance of DINOv2-ViT-giant, DINOv3-ViT-large, RETFound-DINOv2, and RETFound-MAE across the ten tasks was 0.800, 0.816, 0.830, and 0.809, respectively (RETFound-DINOv2 significantly outperformed DINOv3-ViT-large, the second-best performing model). The average ranks of them were 3.2, 2.3, 1.7, and 2.8, respectively, confirming the superior performance of RETFound-DINOv2, followed by DINOv3-ViT-large. A similar trend was observed under linear probe adaptation (Supplementary Figures 8 and 9). Full results are shown in Supplementary Table 3.

We assessed the generalisability of DINOv2-ViT-giant, DINOv3-ViT-large, and RETFound models via external evaluation. For systemic disease prediction, models fine-tuned on AlzEye were evaluated on UK Biobank, which differs in imaging devices and participant demographics. RETFound-DINOv2 showed the strongest external performance, with the highest average AUROC (0.599), compared with RETFound-MAE (0.571), DINOv2-ViT-giant (0.560), and DINOv3-ViT-large (0.549); AUPRC and F1 followed the same ranking (Supplementary Figure 10). For ocular disease detection, AUROC values were similar across models, whereas RETFound-DINOv2 achieved the highest F1 score (0.402), followed by DINOv3-ViT-large (0.381), RETFound-MAE (0.367), and DINOv2-ViT-giant (0.365) (Supplementary Figure 11).

**Fine-tuning versus linear probing**

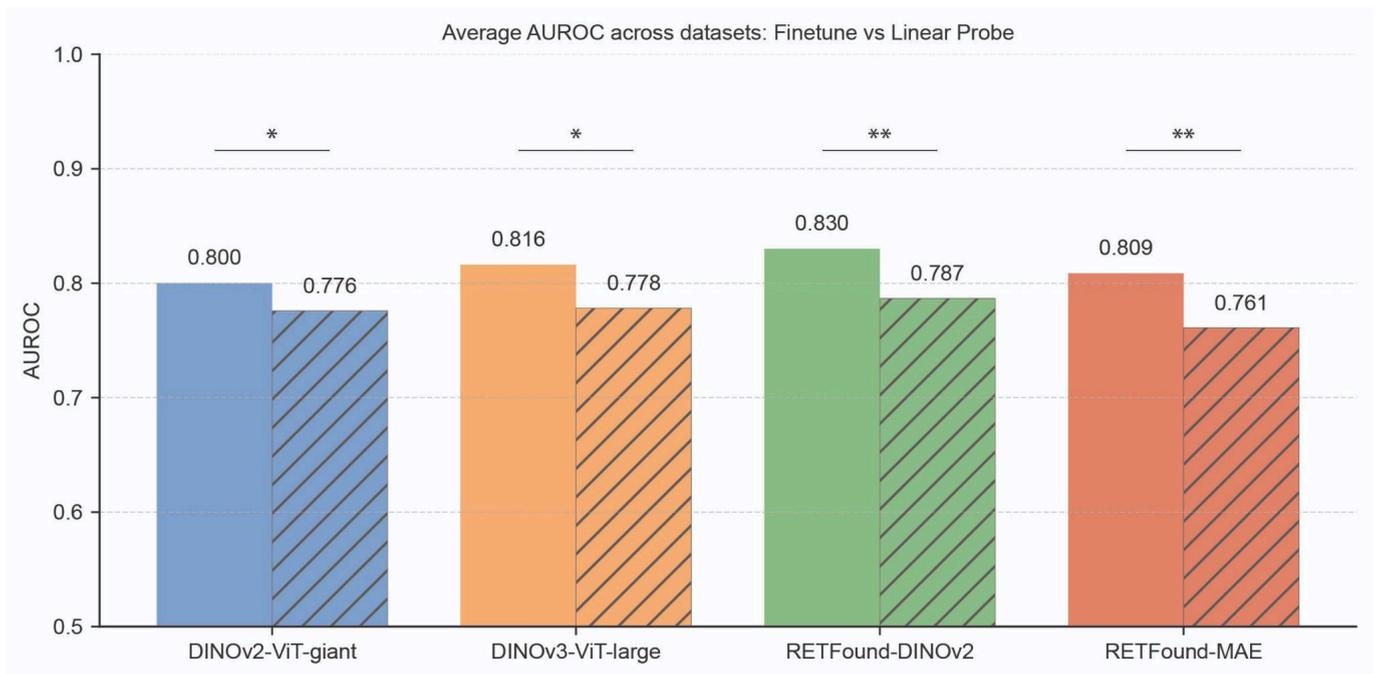

Figure 5. Performance comparison between adaptation methods, fine-tuning (solid bars) and linear probing (hatched bars), for DINOv2-ViT-giant, DINOv3-ViT-large, RETFound-DINOv2, and RETFound-MAE across ten tasks. p-value was calculated using a two-sided Wilcoxon signed-rank test. * indicates p<0.05; ** indicates p<0.01.

Figure 5 shows the average AUROC performance between two adaptation strategies: fine-tuning (solid bars) and linear probing (hatched bars) for DINOv2-ViT-giant, DINOv3-ViT-Large, RETFound-DINOv2 and RETFound-MAE. Fine-tuning, which updates all model parameters, consistently outperforms linear probing, where only a lightweight classifier is trained on frozen features. Average AUROC values range from 0.800 to 0.830 for fine-tuned models compared to 0.761 to 0.787 for linear probing. Among the four models, RETFound-DINOv2 achieves the highest AUROC with fine-tuning (0.830) and linear probing (0.787). Despite the statistically significant performance difference, linear probing still delivered strong performance across all models.

**Data efficiency comparison**

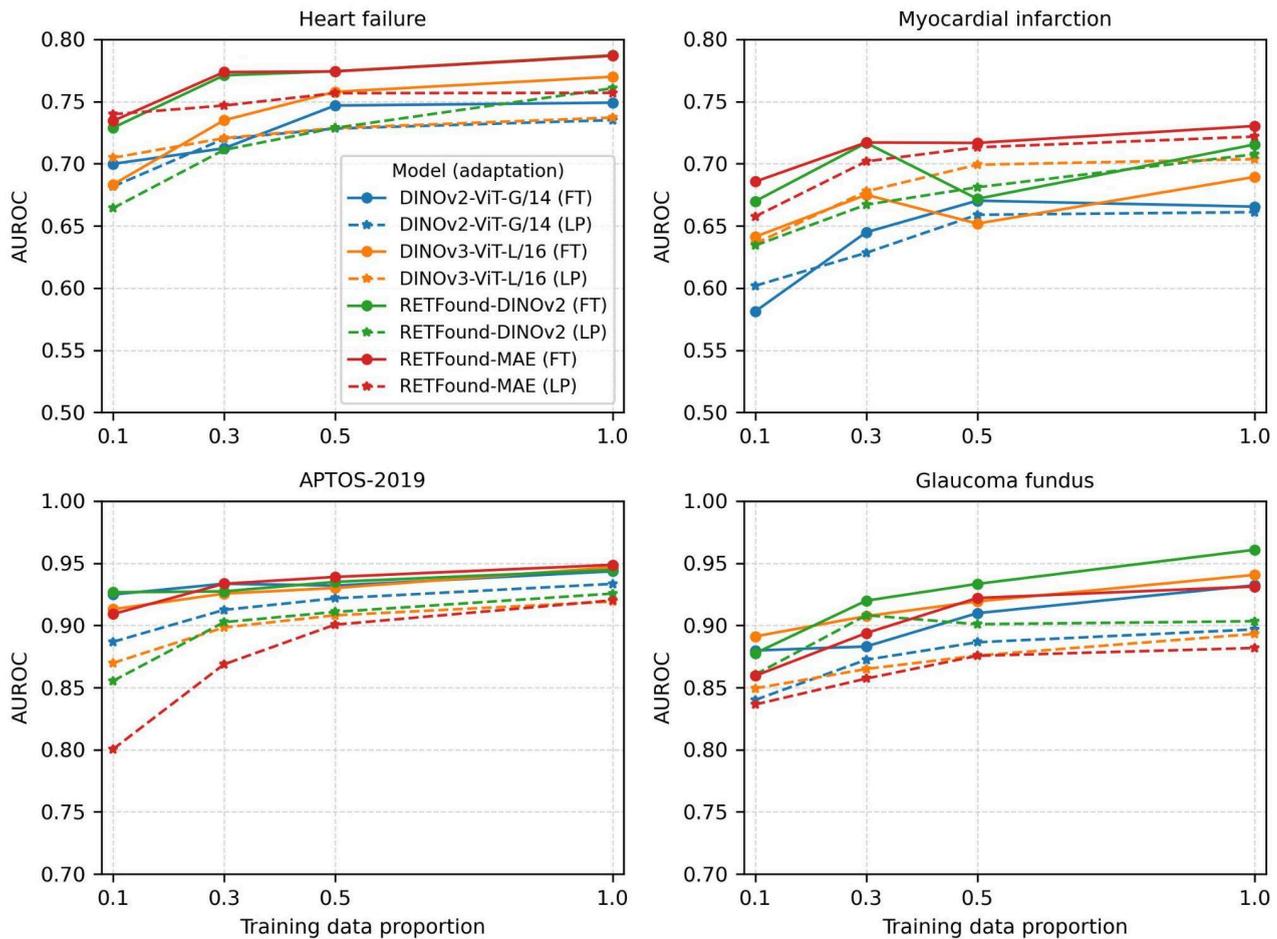

Figure 6. Data efficiency of DINOv2-ViT-Giant, DINOv3-ViT-Large, RETFound-DINOv2, and RETFound-MAE in downstream tasks. Models were trained with varying proportions of the training data and evaluated on the full test sets. Dashed lines indicate linear probing results, while solid lines indicate fine-tuning results. FT represents fine-tuning, and LP represents linear probing.

We evaluated and compared the data efficiency of DINOv2-ViT-giant, DINOv3-ViT-large, RETFound-DINOv2, and RETFound-MAE on downstream tasks. From Figure 6 and Supplementary Figure 12, all four models demonstrated good data efficiency, achieving decent performance with small proportions of training data. For example, RETFound-DINOv2 achieved an AUROC of 0.729 using 10% of training data, corresponding to 92.6% of its full-data performance in heart failure prediction. The performance increased considerably when the training data increased from 10% to 30%, while improvement became more incremental thereafter. With 10% of training data for fine-tuning, DINOv2-ViT-giant, DINOv3-ViT-large, RETFound-DINOv2, and RETFound-MAE achieved average AUROC of 0.705, 0.728, 0.720, and 0.701 across ten tasks. The average rankings among these four models are 2.7, 2.2, 2.2, and 2.9, highlighting the better data efficiency of DINOv3-ViT-large and RETFound-DINOv2.

**Adaptation efficiency comparison**

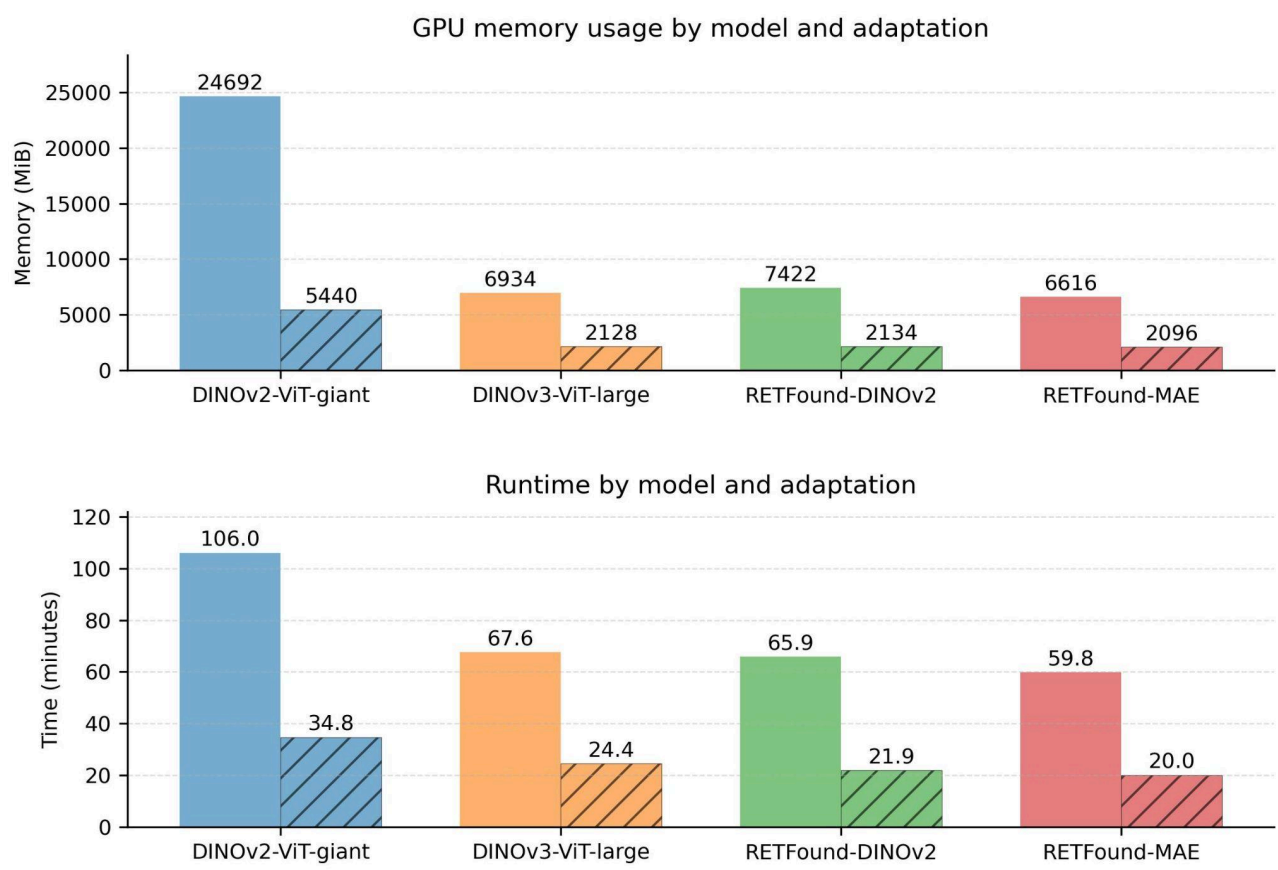

Figure 7. Computational resources and runtime required for adapting DINOv2-ViT-giant, DINOv3-ViT-large, RETFound-DINOv2, and RETFound-MAE to the APTOS-2019 dataset, a diabetic retinopathy dataset including 3,662 images. The testing was conducted on a single NVIDIA A100 80GB GPU. Solid bars indicate fine-tuning efficiency, and hatched bars indicate linear probing efficiency.

Since adaptation efficiency, such as required computational resources and training time, is critical for ensuring broad applicability, we compared the adaptation efficiency of DINOv2-ViT-giant, DINOv3-ViT-large, RETFound-DINOv2, and RETFound-MAE on the APTOS-2019 diabetic retinopathy dataset (3,662 images). Specifically, we recorded GPU memory usage and running time in model adaptation (Figure 7). DINOv3-ViT-large, RETFound-DINOv2, and RETFound-MAE all used the ViT-large backbone and demonstrated similar adaptation efficiency, with varied implementations in code libraries. In contrast, DINOv2-ViT-giant, which uses the larger ViT-giant backbone, required approximately 3.5 times more GPU memory and twice the running time.

**Feature analysis**

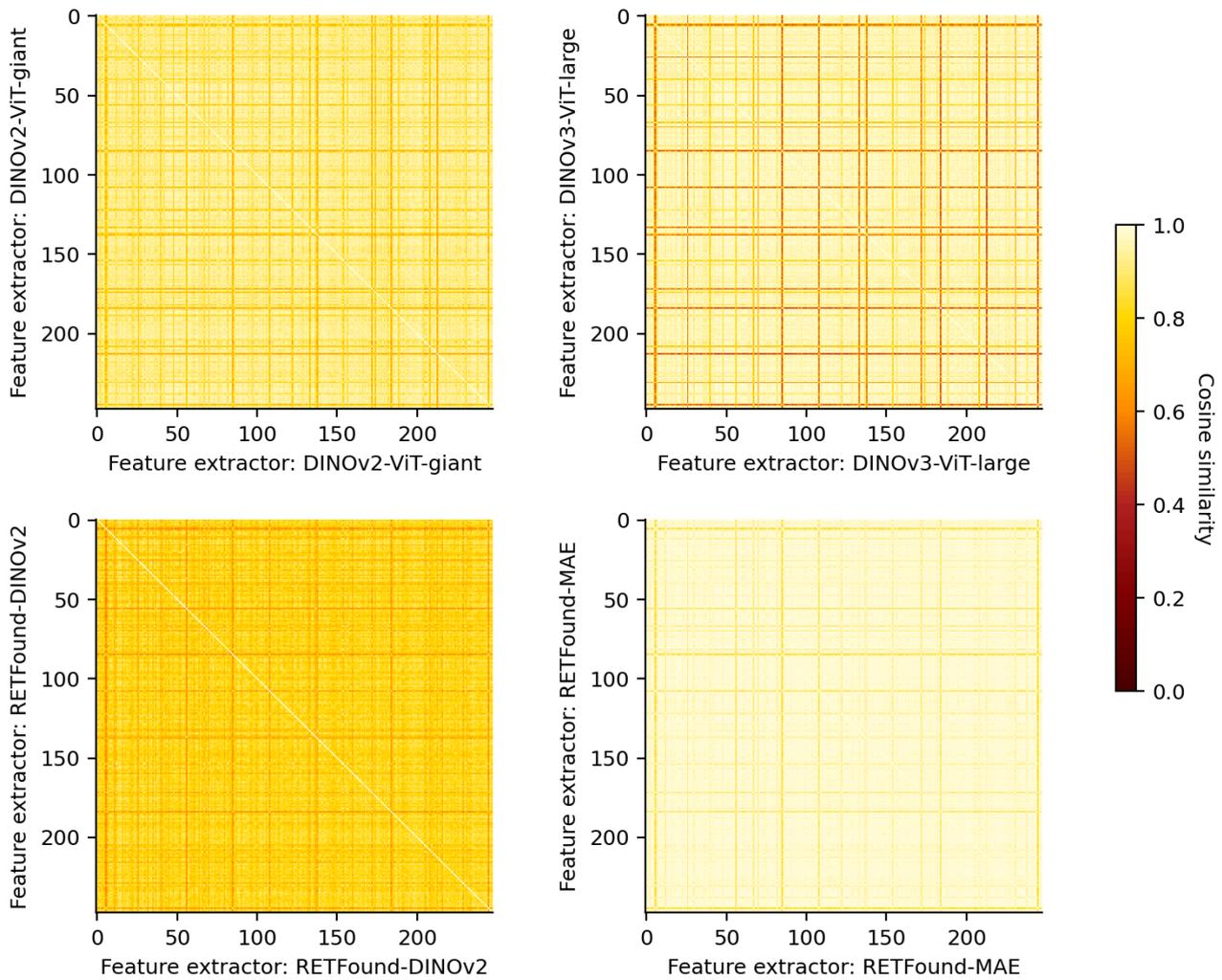

Figure 8. Self-similarity heatmaps of image features extracted by DINOv2-ViT-giant, DINOv3-ViT-large, RETFound-DINOv2, and RETFound-MAE. Features were extracted from 250 retinal images in the UK Biobank, and pairwise similarities were computed to generate a 250×250 matrix for each model. Brighter colours indicate higher similarity. All four models show strong self-similarity across image pairs, with RETFound-MAE exhibiting the highest average similarity (0.975) and RETFound-DINOv2 the lowest (0.798). The X and Y axes represent image IDs.

Figure 8 shows self-similarity heatmaps of image features extracted by four models, DINOv2-ViT-giant, DINOv3-ViT-large, RETFound-DINOv2, and RETFound-MAE, across 250 retinal images randomly sampled from the UK Biobank. Each heatmap depicts the cosine similarity between every pair of image features extracted by the same model, with values closer to 1 (lighter colours) indicating higher similarity. Overall, all models produced features with high similarity. The average pairwise similarity was 0.897 for DINOv2-ViT-giant, 0.915 for DINOv3-ViT-large, 0.798 for RETFound-DINOv2, and 0.975 for RETFound-MAE. Among these, RETFound-DINOv2 extracted more discriminative features with lower similarity, likely contributing to its strong downstream performance by allowing classifiers to better separate subtle clinical signals during fine-tuning or linear probing.

Additionally, we measured cross-model similarity for features extracted by DINOv3-ViT-large, RETFound-DINOv2, and RETFound-MAE (excluding DINOv2-ViT-giant, as it has a different feature size). Cross-model similarity was consistently very low, even for the same image, as indicated by the diagonal lines (Supplementary Figure 13). This suggests that foundation models learn substantially different feature spaces, underscoring the importance of feature alignment when integrating them for clinical applications.

**Discussion**

This study provides systematic comparisons between generalist vision foundation models and specialist retinal foundation models. By benchmarking DINOv2, DINOv3, and RETFound across ten diverse ocular and systemic disease detection tasks, and further examining their data efficiency and adaptation efficiency, we demonstrate both the potential of directly adapting cutting-edge, generalist foundation models to clinical applications and the added value of domain-specific pre-training on retinal images. These findings provide evidence to guide next steps for leveraging foundation models in clinical tasks with retinal images, paving the way for future advancement of retinal foundation models and their deployment.

Our results demonstrate that larger-scale architectures within the DINOv2 series, particularly ViT-giant, consistently achieve superior performance across both ocular and systemic prediction tasks, suggesting that model scale remains a critical factor in transferability to medical imaging. Within the DINOv3 series, ViT-large achieved the best balance of performance across tasks, while ViT-small+ and ViT-huge+ showed limited performance, which differs from the findings reported in natural-image applications [25]. These findings suggest that scaling laws derived from generalist foundation models trained on natural images may not directly translate to medical imaging. Dedicated exploration for clinical applications is therefore essential to identify the optimal balance between model performance and computational resources.

Although DINOv2-ViT-giant contains more parameters than DINOv3-ViT-large, the performance comparison shows that parameter scale alone does not determine downstream effectiveness. DINOv3-ViT-large achieved higher AUROC, AUPRC, and F1 score than DINOv2-ViT-giant under fine-tuning, despite being nearly 3.5 times smaller. This reflects the advantages credited to DINOv3's training paradigm, which benefited from substantially larger pre-training data (~1.7 billion natural images) as well as methodological advances such as Gram Anchoring. These factors likely enabled the DINOv3 models to learn more transferable representations, highlighting the significance of data scale and pre-training strategy for foundation models development and their performance in downstream tasks.

Domain-specific pre-training with RETFound exhibited advantages over generalist DINOv2 and DINOv3 models in certain clinically relevant applications. RETFound-DINOv2 achieved the best average performance over ten tasks, outperforming the most competitive baselines DINOv2-ViT-giant and DINOv3-ViT-large. These results confirm that domain-specific pre-training remains highly valuable. The lower pairwise similarity of features extracted by RETFound-DINOv2 indicates that its embeddings are more discriminative and less redundant compared to the other models. This likely contributes to its strong downstream performance, as more distinguishable features make it easier for classifiers to separate subtle clinical signals during fine-tuning or linear probing. When comparing RETFound-DINOv2 and RETFound-MAE, we observe that RETFound-DINOv2 generally achieved a higher performance.

Both models were pre-trained on the same data with identical computational resources, though their reported results may not fully reflect optimised performance. These results highlight RETFound-DINOv2 as a strong first-choice foundation model for downstream applications in both ocular and systemic disease prediction. More importantly, the superior performance of RETFound-DINOv2 highlights the benefits of pre-training on medical images, even as the latest generalist foundation models continue to show improved transferability.

Our comparisons between fine-tuning and linear probing highlight the importance of the adaptation strategy for clinical deployment. While fine-tuning consistently achieved superior performance, linear probing still delivered competitive results, suggesting that foundation models encode broadly transferable medical features. This is practically significant as linear probing can provide a cost-effective option in resource-constrained settings. Specifically, fine-tuning generally requires several times more memory and runtime than linear probing (Figure 7), highlighting a clear trade-off between performance and efficiency. For instance, fine-tuning DINOv2-ViT-giant requires ~5 times more memory and ~3 times longer runtime than linear probing, translating into a substantial increase in overall cost. Hence, the choice between fine-tuning and linear probing should be informed not only by performance objectives but also by the practical constraints (e.g. computational resources) of healthcare environments, as well as broader societal considerations such as carbon footprints.

DINOv2-ViT-giant, DINOv3-ViT-large, and RETFound demonstrated strong data efficiency when adapted to clinical downstream tasks. With only 10% of the training data, these models retained over 90% of their full-data performance, indicating that foundation models remain effective even in low-data regimes. This is particularly valuable for medical applications where annotated datasets are often limited, as it reduces reliance on large-scale data curation and annotation and shortens training time. Beyond data efficiency, feature similarity analyses offered additional insights into their feature properties. As shown in Figure 8, all models produced highly self-consistent features, albeit with varying levels of granularity. Cross-model comparisons (Supplementary Figure 11) further revealed minimal similarity between models. This divergence suggests that different foundation models have distinct feature spaces, highlighting that applying them to a curated library of clinically relevant tasks remains an active area of exploration. Regarding foundation model development, the limited cross-model feature similarity underscores the importance of feature alignment when developing multimodal models or integrating multiple foundation models.

Despite offering systematic evaluations, our study has a few limitations. First, the analysis was limited to vision foundation models, whereas future work should extend to multimodal models, such as vision-language models (e.g. PaliGemma [37] and Qwen [38]), to assess the transferability of broader generalist models in clinical applications. Second, medical foundation models developed based on DINOv3 were not included in our evaluation due to the recent release of DINOv3. Incorporating this model would help determine whether further gains can be achieved by directly adapting state-of-the-art generalist foundation models to the medical domain. Additionally, benchmarking the models on other major categories of tasks, such as segmentation [39] and regression, would help understand their all-around capabilities. Third, we did not evaluate DINOv3-ViT-7B, as it requires substantially higher computational and data requirements for both fine-tuning and linear probing, which would fall outside the scope of our main focus on efficient adaptation for clinical applications. While we did not investigate the underlying reasons for the lower performance of DINOv3-ViT-small+ and DINOv3-ViT-huge+, we provide a thorough assessment of their transferability to retinal imaging. Fourth, examining how the studied foundation models perform across varied image sizes would help improve their adaptation efficiency. Finally, our downstream tasks focused on eye disease detection and systemic disease prediction using retinal images. Extending the analysis to other medical imaging modalities could provide additional insights into the efficiency and adaptability of generalist foundation models in clinical applications.

In summary, our study evaluates and compares generalist and specialist retinal foundation models in detecting eye diseases and predicting systemic diseases. Scaling generalist foundation models, as seen with DINOv2 and DINOv3, provides strong adaptability across diverse tasks, while specialist models like RETFound-DINOv2 offer clear advantages in clinical downstream applications. Combining the scalability and broad transferability of new and advanced generalist models with the medical specificity of domain-pretrained models may represent an effective pathway toward medical foundation models that are both computationally efficient and clinically impactful.

**Data Availability**

The AlzEye data consists of routinely collected healthcare data. Due to their sensitive nature, the dataset is subject to controlled access through a structured application process. The AlzEye dataset is subject to the contractual restrictions of the data sharing agreements between National Health Service Digital, Moorfields Eye Hospital and University College London, and is not available for access beyond the AlzEye research team. National and international collaborations are welcomed, although restrictions on access to the cohort mean that only the AlzEye researchers can directly analyse individual-level systemic health data. Interested collaborators should contact P.A.K.

Data for ocular disease experiments are publicly available online and can be accessed through the following links: IDRiD (https://ieee-dataport.org/open-access/indian-diabetic-retinopathy-image-dataset-idrid), MESSIDOR2 (https://www.adcis.net/en/third-party/messidor2/), APTOS2019 (https://www.kaggle.com/competitions/aptos2019-blindness-detection/data), Glaucoma Fundus (https://dataverse.harvard.edu/dataset.xhtml?persistentId=doi:10.7910/DVN/1YRRAC), PAPILA (https://figshare.com/articles/dataset/PAPILA/14798004/1), and Retina (https://www.kaggle.com/datasets/jr2ngb/cataractdataset).

**Code Availability**

The code used to fine-tune and evaluate DINOv3, DINOv2, and RETFound from Y.Z. is available at https://github.com/rmaphoh/RETFound, which is based on PyTorch. All RETFound model weights are available at https://huggingface.co/YukunZhou. DINOv3 model weights are available upon application at https://github.com/facebookresearch/dinov3. DINOv2 model weights are available at https://github.com/facebookresearch/dinov2. Images were processed with the automated retinal image analysis tool AutoMorph v.1.0 (https://github.com/rmaphoh/AutoMorph). Results were further analysed and visualised with Python v.3.11.0, NumPy v.1.26.4, SciPy v.1.15.2, Matplotlib v.3.8.4, pandas v.1.5.0, Scikit-Learn v.1.4.2 and Pillow v.10.2.0.

**Ethics statement**

This study involves human participants and was approved by the London-Central Research Ethics Committee (18/LO/1163, approved 1 August 2018), Advanced statistical modelling of multimodal data of genetic and acquired retinal diseases (20/HRA/2158, approved 5 May 2020), and Confidential Advisory Group for Section 251 support (18/CAG/0111, approved 13 September 2018). The National Health Service Health Research Authority gave final approval on 13 September 2018. Moorfields Eye Hospital

NHS Foundation Trust validated the de-identifications for MEH data. Only de-identified retrospective data were used for research.

**Acknowledgment**

Y.Z. is supported by Wellcome Trust Early-career Award (318987/Z/24/Z). Y.C.T. is supported by the National Medical Research Council of Singapore (NMRC/MOH/ HCSAINV21nov-0001). C.Y.C. is supported by the Research Grants Council Hong Kong (General Research Fund, ref.14101324). D.C.A is supported by Engineering and Physical Sciences Research Council EP/M020533/1, EP/R014019/1 and EP/V034537/1. P.A.K. is supported by a UK Research & Innovation Future Leaders Fellowship (MR/T019050/1), the Moorfields Eye Charity with The Rubin Foundation Charitable Trust (GR001753), and an Alcon Research Institute Senior Investigator Award. We acknowledge the computational resources supported by the UCL Computer Science Cluster and the Advanced Research Computing UAI platform.

**References**

1. Bommasani R, Hudson DA, Adeli E, Altman R, Arora S, von Arx S, et al. On the Opportunities and Risks of Foundation Models. arXiv [cs.LG]. 2021. Available: http://arxiv.org/abs/2108.07258

2. Wiggins WF, Tejani AS. On the Opportunities and Risks of Foundation Models for Natural Language Processing in Radiology. Radiol Artif Intell. 2022;4: e220119.

3. Moor M, Banerjee O, Abad ZSH, Krumholz HM, Leskovec J, Topol EJ, et al. Foundation models for generalist medical artificial intelligence. Nature. 2023;616: 259–265.

4. He Y, Huang F, Jiang X, Nie Y, Wang M, Wang J, et al. Foundation Model for Advancing Healthcare: Challenges, Opportunities and Future Directions. IEEE Rev Biomed Eng. 2025;18: 172–191.

5. Zhang S, Metaxas D. On the challenges and perspectives of foundation models for medical image analysis. Med Image Anal. 2024;91: 102996.

6. Chia MA, Zhou Y, Keane PA. A new foundation model for multimodal ophthalmic images: Advancing disease detection and prediction. NEJM AI. 2024;1. doi:10.1056/aie2401024

7. Zhou Y, Chia MA, Wagner SK, Ayhan MS, Williamson DJ, Struyven RR, et al. A foundation model for generalizable disease detection from retinal images. Nature. 2023;622: 156–163.

8. Huang D-S, Chen H, Li B, Zhang Q. Advanced Intelligent Computing Technology and Applications: 21st International Conference, ICIC 2025, Ningbo, China, July 26–29, 2025, Proceedings, Part VIII. Springer Nature; 2025.

9. Li J, Guan Z, Wang J, Cheung CY, Zheng Y, Lim L-L, et al. Integrated image-based deep learning and language models for primary diabetes care. Nat Med. 2024;30: 2886–2896.

10. Tiu E, Talius E, Patel P, Langlotz CP, Ng AY, Rajpurkar P. Expert-level detection of pathologies from unannotated chest X-ray images via self-supervised learning. Nat Biomed Eng. 2022. doi:10.1038/s41551-022-00936-9

11. Pai S, Bontempi D, Hadzic I, Prudente V, Sokač M, Chaunzwa TL, et al. Foundation model for cancer imaging biomarkers. Nat Mach Intell. 2024;6: 354–367.

12. Tanno R, Barrett DGT, Sellergren A, Ghaisas S, Dathathri S, See A, et al. Collaboration between clinicians and vision-language models in radiology report generation. Nat Med. 2025;31: 599–608.

13. Huang Z, Bianchi F, Yuksekgonul M, Montine TJ, Zou J. A visual-language foundation model for pathology image analysis using medical Twitter. Nat Med. 2023;29: 2307–2316.

14. Lu MY, Chen B, Williamson DFK, Chen RJ, Liang I, Ding T, et al. A visual-language foundation model for computational pathology. Nat Med. 2024;30: 863–874.

15. Xu H, Usuyama N, Bagga J, Zhang S, Rao R, Naumann T, et al. A whole-slide foundation model for digital pathology from real-world data. Nature. 2024;630: 181–188.

16. He K, Fan H, Wu Y, Xie S, Girshick R. Momentum contrast for unsupervised visual representation learning. 2020 IEEE/CVF Conference on Computer Vision and Pattern Recognition (CVPR). IEEE; 2020. pp. 9729–9738.

17. Caron M, Touvron H, Misra I, Jégou H, Mairal J, Bojanowski P, et al. Emerging properties in self-supervised vision transformers. arXiv [cs.CV]. 2021. pp. 9650–9660. Available: http://openaccess.thecvf.com/content/ICCV2021/html/Caron_Emerging_Properties_in_Self-Supervised_Vision_Transformers_ICCV_2021_paper.html

18. Oquab M, Darcet T, Moutakanni T, Vo H, Szafraniec M, Khalidov V, et al. DINOv2: Learning robust

visual features without supervision. arXiv [cs.CV]. 2023. Available: http://arxiv.org/abs/2304.07193

19. He K, Chen X, Xie S, Li Y, Doll'ar P, Girshick RB. Masked Autoencoders Are Scalable Vision Learners. Proc IEEE Comput Soc Conf Comput Vis Pattern Recognit. 2021; 15979–15988.

20. Dosovitskiy A, Beyer L, Kolesnikov A, Weissenborn D, Zhai X, Unterthiner T, et al. An image is worth 16x16 words: Transformers for image recognition at scale. Int Conf Learn Represent. 2020;abs/2010.11929. doi:10.11929/1000

21. Singhal K, Tu T, Gottweis J, Sayres R, Wulczyn E, Amin M, et al. Toward expert-level medical question answering with large language models. News@nat,Com. 2025. doi:10.1038/s41591-024-03423-7

22. Zoellin J, Merk C, Buob M, Saad A, Giesser S, Spitznagel T, et al. Block expanded DINORET: Adapting natural domain foundation models for retinal imaging without catastrophic forgetting. arXiv [cs.CV]. 2024. Available: http://arxiv.org/abs/2409.17332

23. Zou K, Goh JHL, Zhou Y, Lin T, Yew SME, Srinivasan S, et al. FusionFM: Fusing eye-specific foundational models for optimized ophthalmic diagnosis. arXiv [cs.CV]. 2025. Available: http://arxiv.org/abs/2508.11721

24. Zhou Y, Wang Z, Wu Y, Ong AY, Wagner S, Ruffell E. Revealing the Impact of Pre-training Data on Medical Foundation Models. 2025. Available: https://www.researchsquare.com/article/rs-6080254/latest

25. Siméoni O, Vo HV, Seitzer M, Baldassarre F, Oquab M, Jose C, et al. DINOv3. arXiv [cs.CV]. 2025. Available: http://arxiv.org/abs/2508.10104

26. Hou Q, Zhou Y, Goh JHL, Zou K, Yew SME, Srinivasan S, et al. Can a natural image-based foundation model outperform a retina-specific model in detecting ocular and systemic diseases? Ophthalmol Sci. 2025; 100923.

27. Wagner SK, Fu DJ, Faes L, Liu X, Huemer J, Khalid H, et al. Insights into Systemic Disease through Retinal Imaging-Based Oculomics. Transl Vis Sci Technol. 2020;9: 6.

28. Zhu Z, Wang Y, Qi Z, Hu W, Zhang X, Wagner SK, et al. Oculomics: Current concepts and evidence. Prog Retin Eye Res. 2025;106: 101350.

29. Wagner S, Hughes F, Cortina-Borja M, Pontikos N, Struyven R, Liu X, et al. AlzEye: longitudinal record-level linkage of ophthalmic imaging and hospital admissions of 353 157 patients in London, UK. BMJ Open. 2022;12. doi:10.1136/bmjopen-2021-058552

30. Porwal P, Pachade S, Kokare M, Deshmukh G, Son J, Bae W, et al. Idrid: Diabetic retinopathy--segmentation and grading challenge. Med Image Anal. 2020;59: 101561.

31. Abràmoff, ; Folk MD, ; Han JC, ; Walker DP, ; Williams JD, ; Russell DF, et al. Automated Analysis of Retinal Images for Detection of Referable Diabetic Retinopathy. JAMA Ophthalmol. 2013;131: 351–357.

32. Decencière E, Zhang X, Cazuguel G, Lay B, Cochener B, Trone C, et al. Feedback on a publicly distributed image database: The Messidor database. Image Anal Stereol. 2014;33: 231.

33. Kovalyk O, Morales-Sánchez J, Verdú-Monedero R, Sellés-Navarro I, Palazón-Cabanes A, Sancho-Gómez J-L. PAPILA: Dataset with fundus images and clinical data of both eyes of the same patient for glaucoma assessment. Sci Data. 2022;9: 291.

34. Ahn JM, Kim S, Ahn K-S, Cho S-H, Lee KB, Kim US. Correction: A deep learning model for the detection of both advanced and early glaucoma using fundus photography. PLoS One. 2019;14: e0211579.

35. Bycroft C, Freeman C, Petkova D, Band G, Elliott LT, Sharp K, et al. The UK Biobank resource with deep phenotyping and genomic data. Nature. 2018;562: 203–209.

36. Zhou Y, Wagner SK, Chia MA, Zhao A, Woodward-Court P, Xu M, et al. AutoMorph: Automated Retinal Vascular Morphology Quantification Via a Deep Learning Pipeline. Transl Vis Sci Technol. 2022;11: 12.

37. Beyer L, Steiner A, Pinto AS, Kolesnikov A, Wang X, Salz D, et al. PaliGemma: A versatile 3B VLM for transfer. arXiv [cs.CV]. 2024. Available: http://arxiv.org/abs/2407.07726

38. Bai J, Bai S, Chu Y, Cui Z, Dang K, Deng X, et al. Qwen Technical Report. arXiv [cs.CL]. 2023. Available: http://arxiv.org/abs/2309.16609

39. Gao Y, Li H, Yuan F, Wang X, Gao X. Dino U-Net: Exploiting high-fidelity dense features from foundation models for medical image segmentation. arXiv [cs.CV]. 2025. Available: http://arxiv.org/abs/2508.20909

**Supplementary materials**

**Supplementary Tables:**
https://docs.google.com/spreadsheets/d/1eRD44imIhicHehRrP7w7LBSHj8ztLEFv/edit?usp=sharing&ouid=115097033631735657188&rtpof=true&sd=true

**Supplementary Figures:**

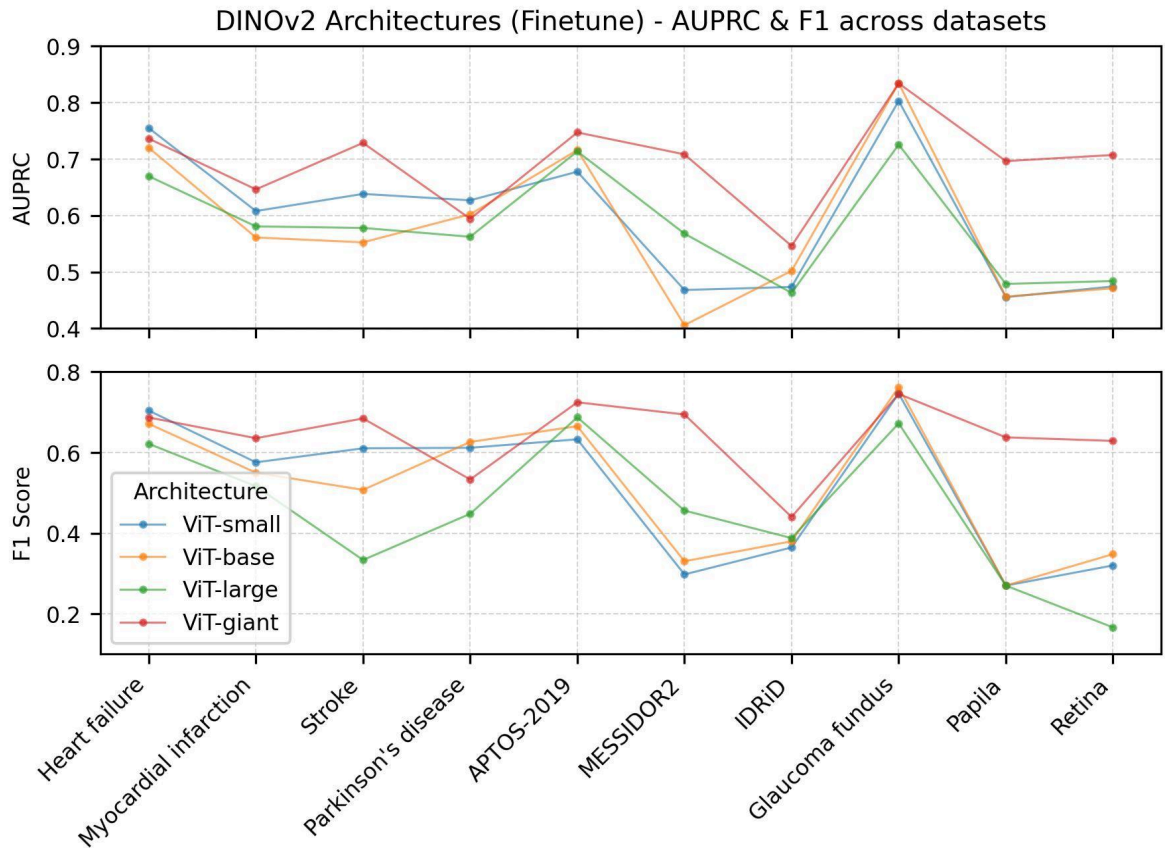

Supplementary Figure 1. Fine-tuning performance comparison of four DINOv2 models. The x-axis denotes the tasks, and the evaluation metrics include AUPRC and F1 score. The average AUPRC of ViT-small, ViT-base, ViT-large, and ViT-giant across ten tasks were 0.598, 0.582, 0.582, and 0.694 respectively. The average rankings of the four models were 2.6, 2.9, 3.1, and 1.4. For the F1 score, the average values were 0.513, 0.511, 0.456, and 0.641. The average rankings of the F1 score were 2.6, 2.4, 3.2, and 1.5 respectively.

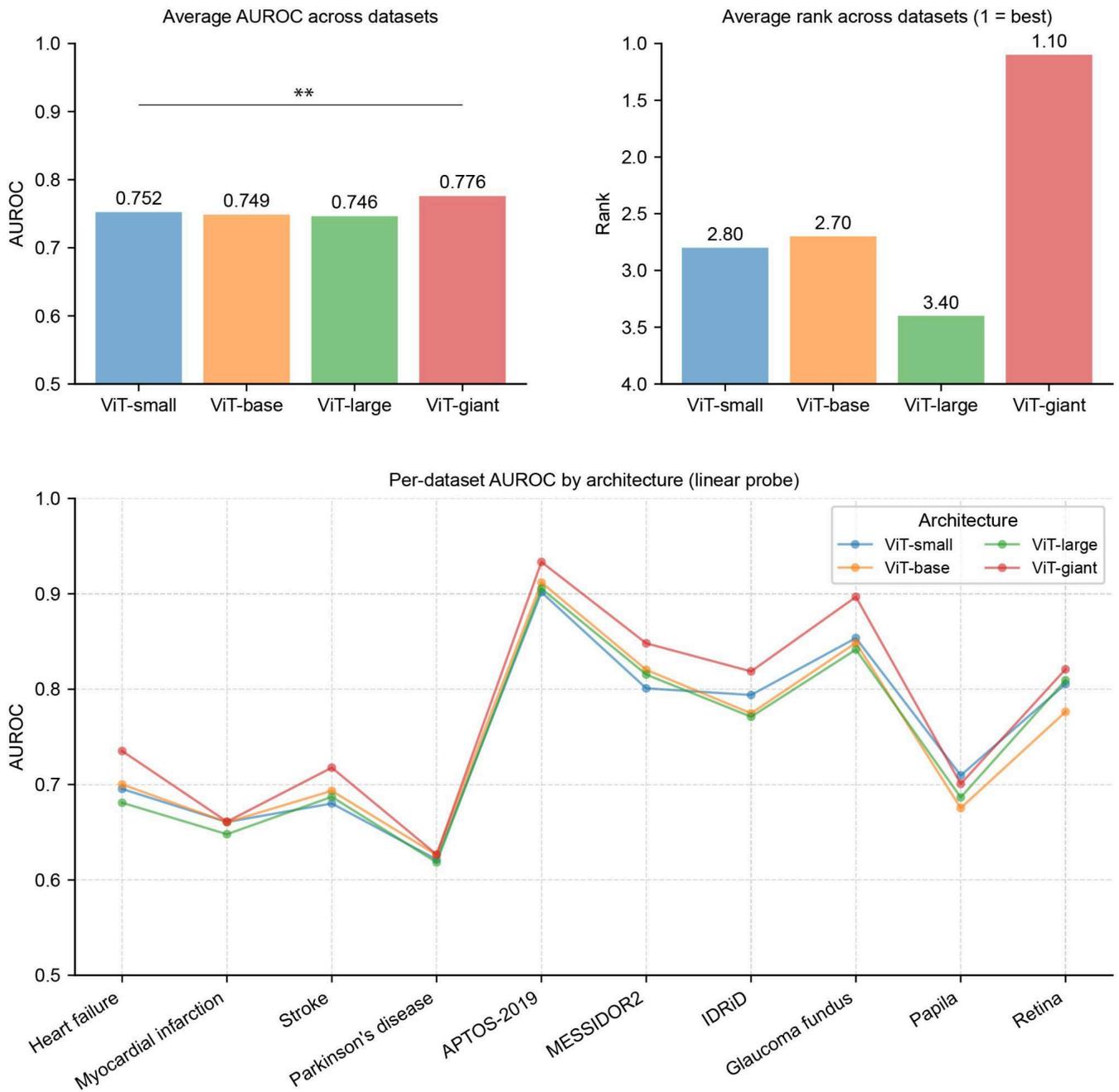

Supplementary Figure 2. Linear probing performance comparison of four DINOv2 models. The x-axis denotes the tasks, and the evaluation metrics include AUROC. The average AUROC of ViT-small, ViT-base, ViT-large, and ViT-giant across ten tasks are 0.752, 0.749, 0.746, and 0.776 respectively. The average rankings of the four models are 2.8, 2.7, 3.4, and 1.1. ViT-giant significantly outperformed ViT-small (p<0.01), the second-best performing model.

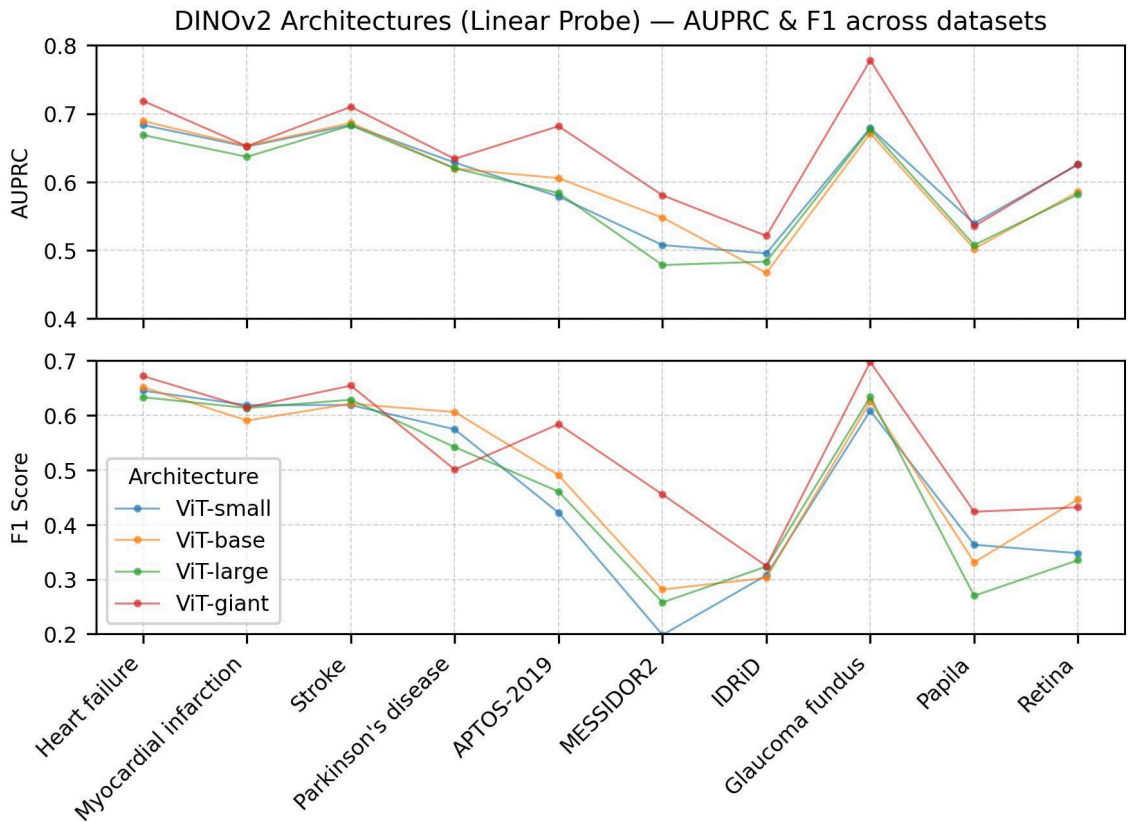

Supplementary Figure 3. Linear probing performance comparison of four DINOv2 models. The x-axis denotes the tasks, and the evaluation metrics include AUPRC and F1 score. The average AUPRC of ViT-small, ViT-base, ViT-large, and ViT-giant across ten tasks were 0.607, 0.603, 0.592, and 0.644 respectively. The average rankings of the four models were 2.5, 2.8, 3.5, and 1.2. For the F1 score, the average values were 0.471, 0.495, 0.470, and 0.536. The average rankings of the F1 score were 3.0, 2.5, 3.0, and 1.5 respectively.

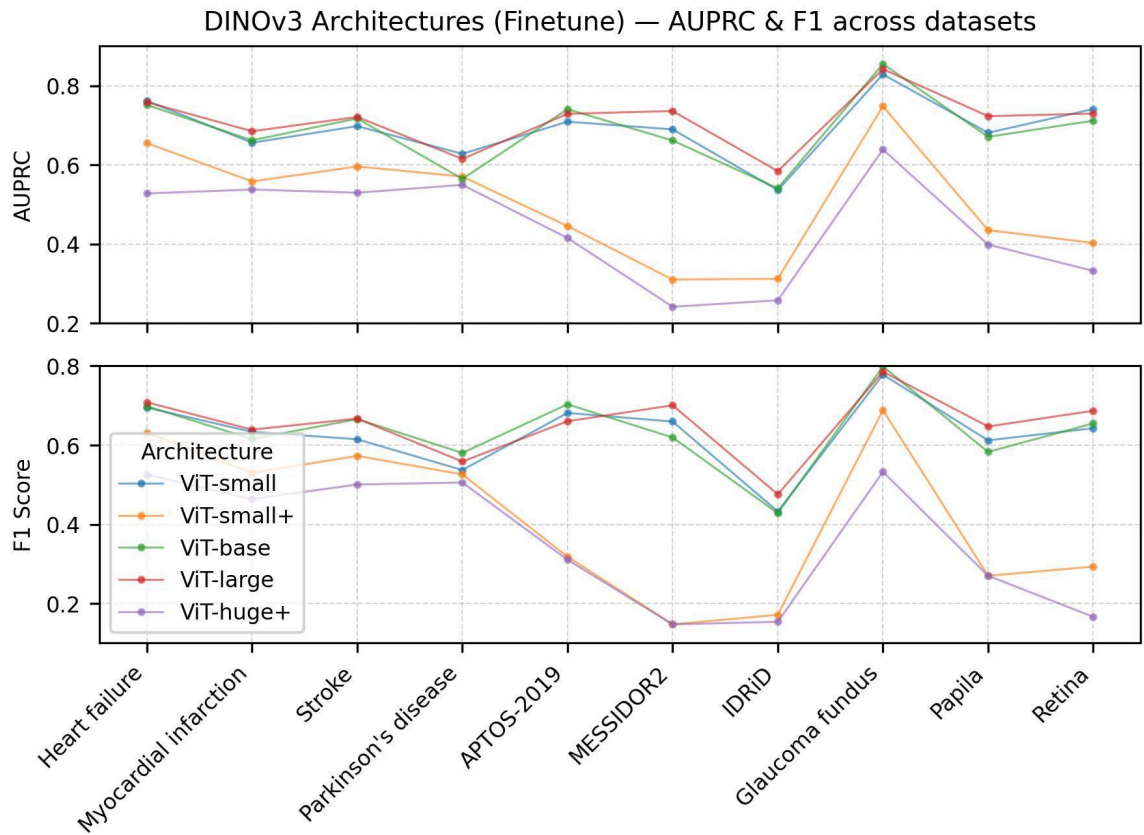

Supplementary Figure 4. Fine-tuning performance comparison of five DINOv3 models. The x-axis denotes the tasks, and the evaluation metrics include AUPRC and F1 score. The average AUPRC of ViT-small, ViT-small+, ViT-base, ViT-large, and ViT-huge+ across ten tasks were 0.693, 0.503, 0.688, 0.712, and 0.443 respectively. The average rankings of the four models were 2.2, 3.9, 2.4, 1.5, and 5.0. For the F1 score, the average values were 0.629, 0.415, 0.635, 0.653, and 0.358. The average rankings of the F1 score were 2.5, 4.0, 2.1, 1.4, and 4.8 respectively.

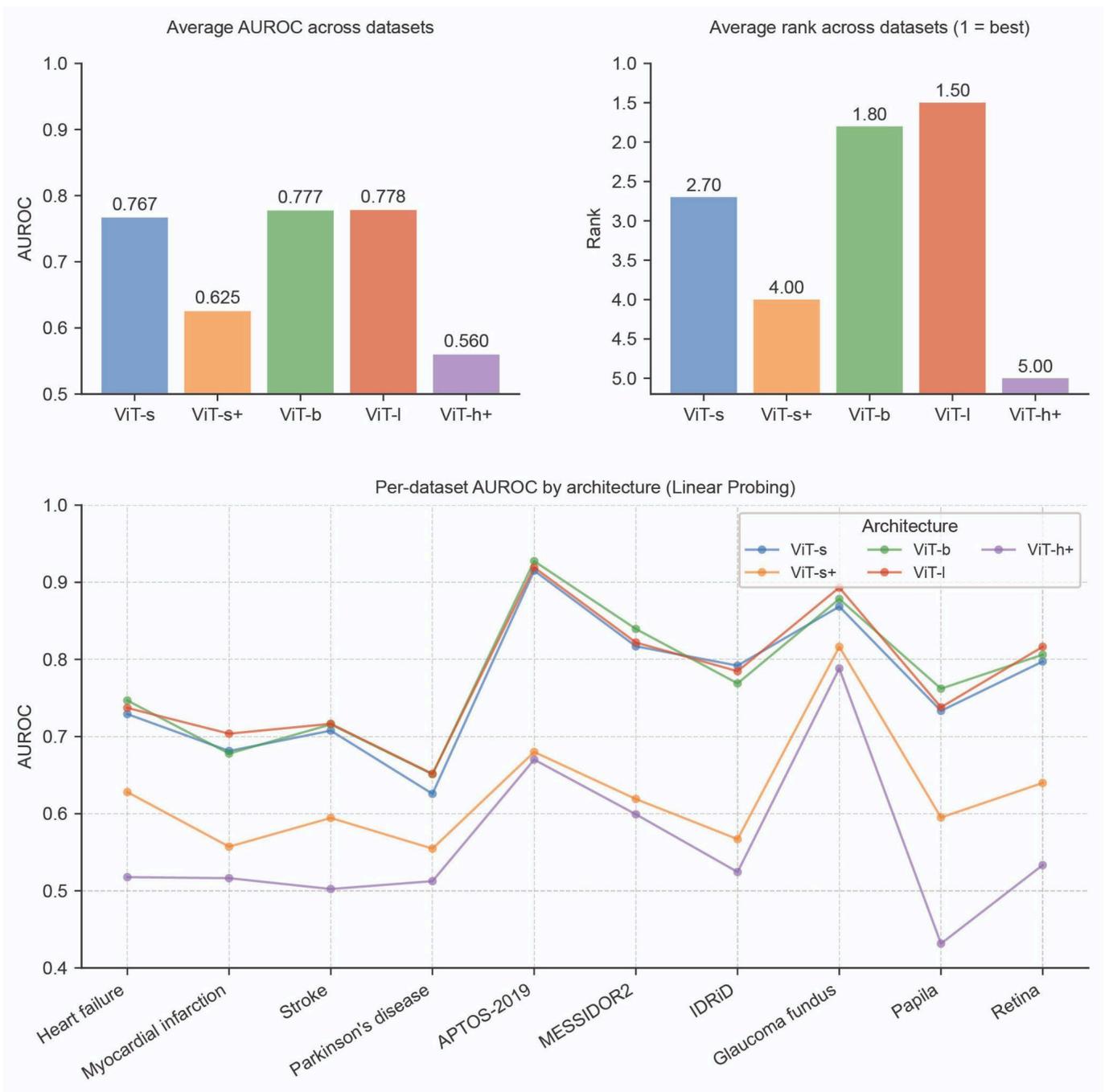

Supplementary Figure 5. Linear probing performance comparison of five DINOv3 models. The x-axis denotes the tasks, and the evaluation metrics include AUROC. The average AUROC of the five DINOv3 models across ten tasks were 0.767, 0.625, 0.777, 0.778, and 0.560 respectively. The average rankings of ViT-small, ViT-small+, ViT-base, ViT-large, and ViT-huge+ across the ten tasks were 2.7, 4.0, 1.8, 1.5, and 5.0 respectively.

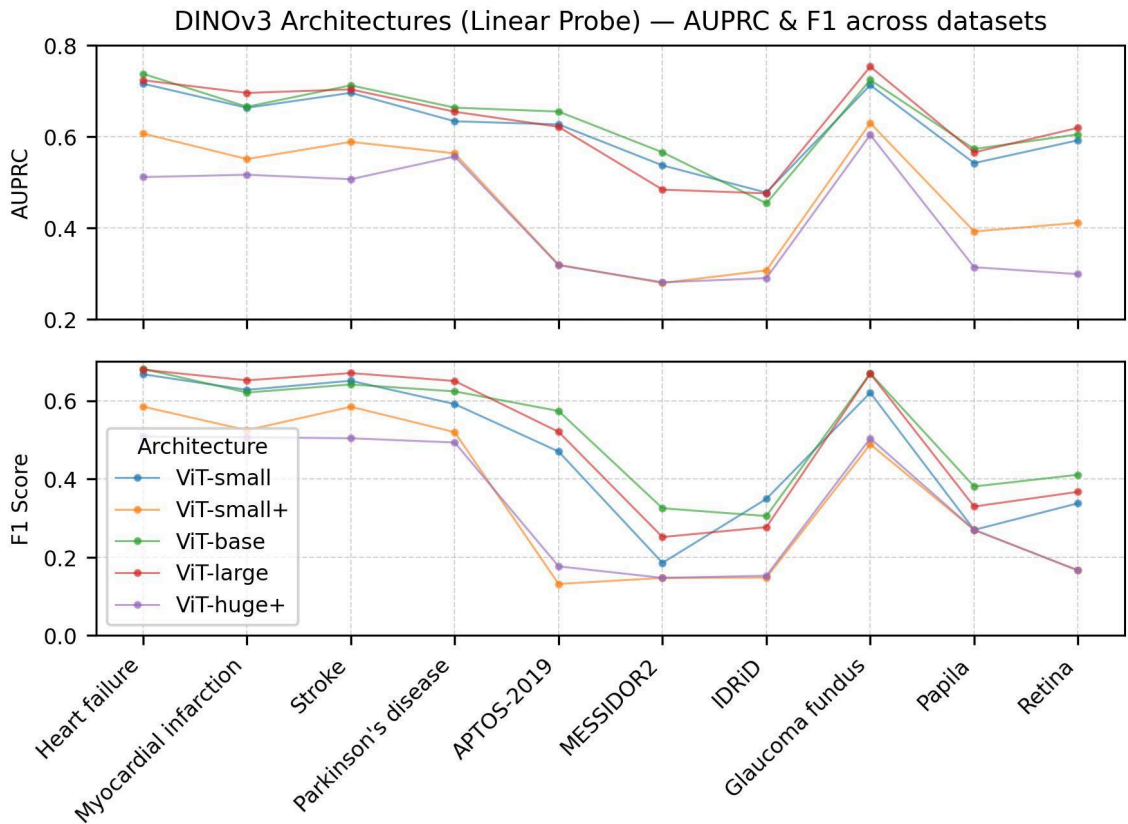

Supplementary Figure 6. Linear probing performance comparison of five DINOv3 models. The x-axis denotes the tasks, and the evaluation metrics include AUPRC and F1 score. The average AUPRC of ViT-small, ViT-small+, ViT-base, ViT-large, and ViT-huge+ across ten tasks were 0.620, 0.465, 0.636, 0.630, and 0.420 respectively. The average rankings of the four models were 2.6, 4.1, 1.5, 1.9, and 4.9. For the F1 score, the average values were 0.477, 0.357, 0.523, 0.507, and 0.343. The average rankings of the F1 score were 2.6, 4.3, 1.6, 1.8, and 4.3 respectively.

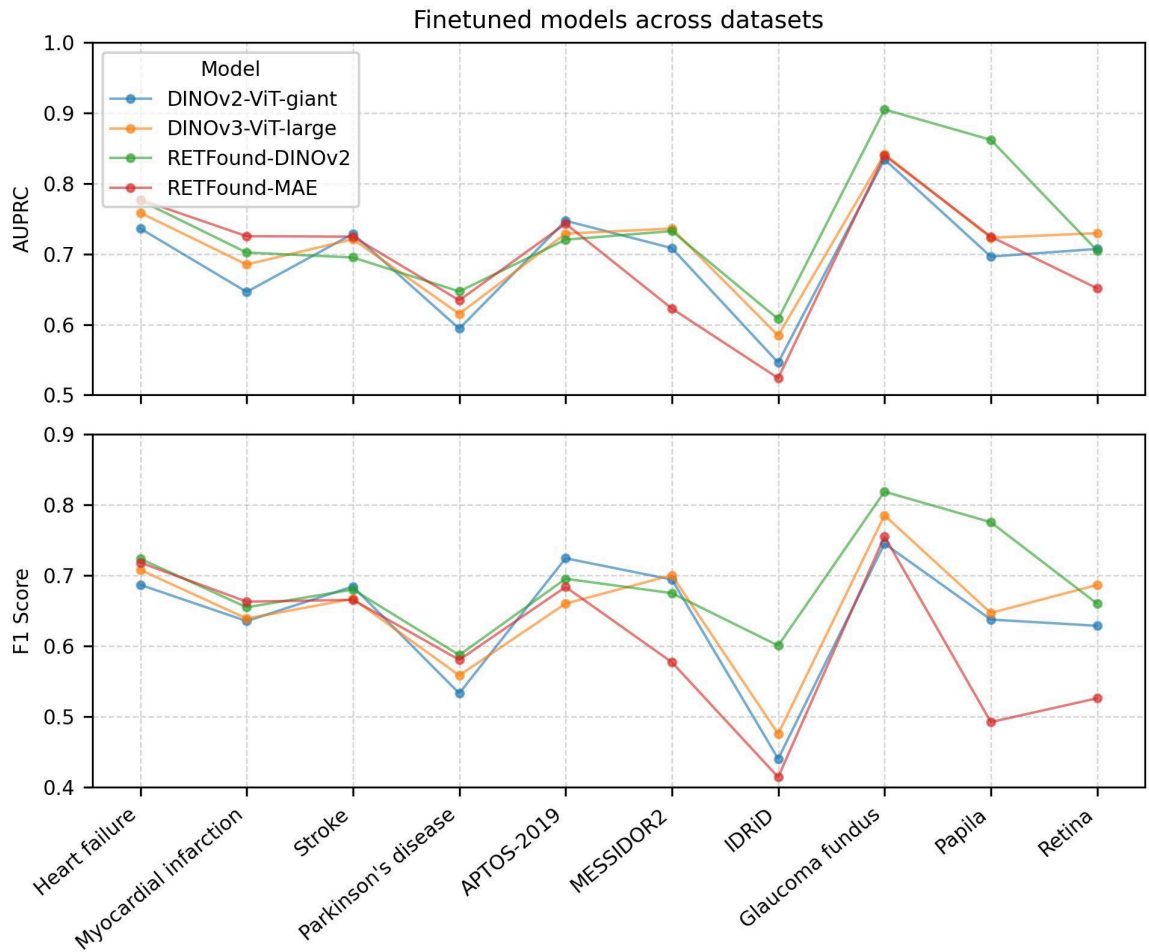

Supplementary Figure 7. Fine-tuning performance comparison of DINOv2-ViT-giant, DINOv3-ViT-large, RETFound-DINOv2, and RETFound-MAE. The x-axis denotes the tasks, and the evaluation metrics include AUPRC and F1 score. The average AUPRC of DINOv2-ViT-giant, DINOv3-ViT-large, RETFound-DINOv2, and RETFound-MAE across ten tasks were 0.694, 0.712, 0.735, and 0.697 respectively. The average rankings of the four models were 3.0, 2.4, 2.1, and 2.5. For the F1 score, the average values were 0.641, 0.653, 0.687, and 0.608. The average rankings of the F1 score were 2.9, 2.4, 1.6, and 3.1 respectively.

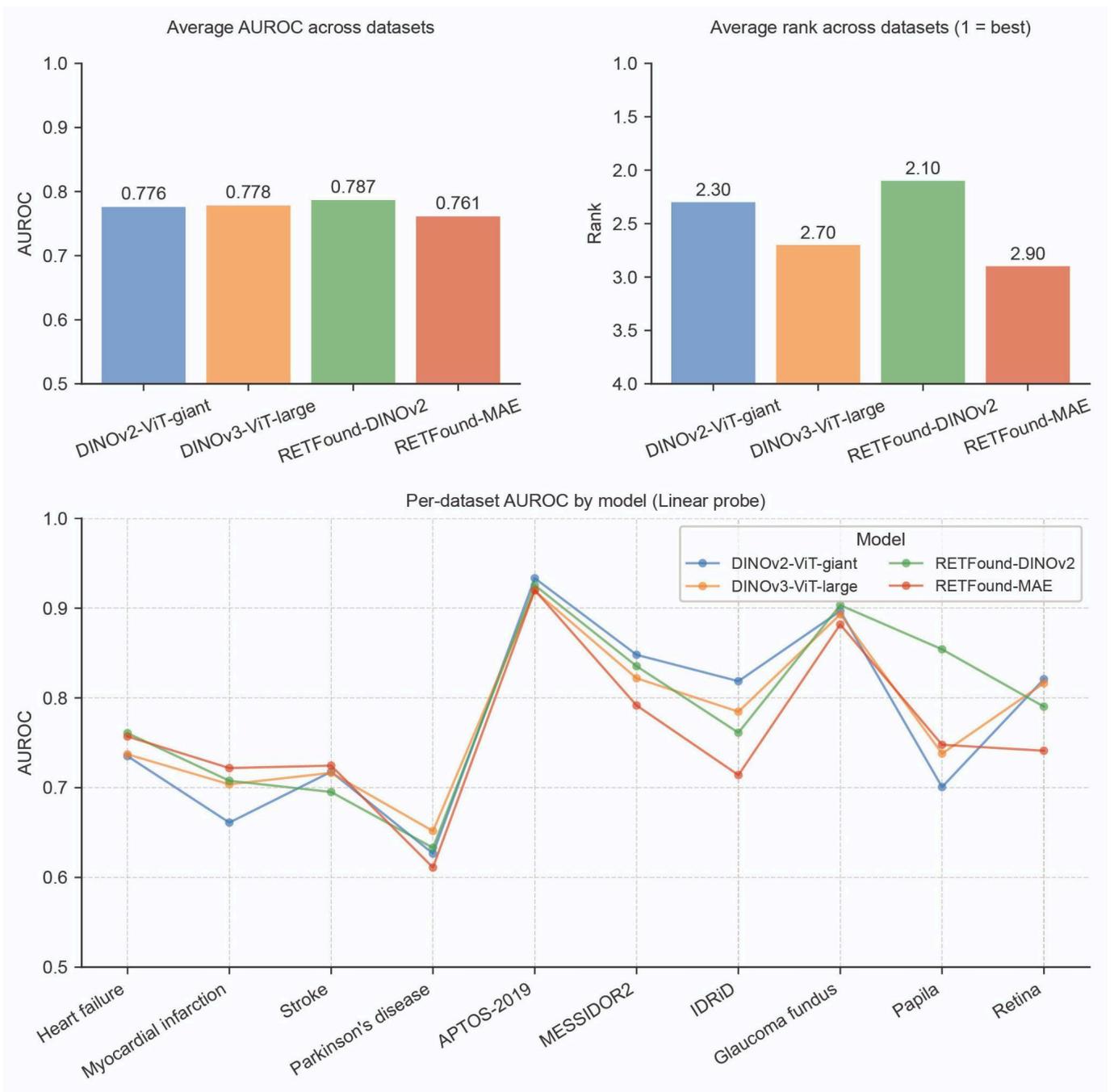

Supplementary Figure 8. Linear probing performance comparison of DINOv2-ViT-giant, DINOv3-ViT-large, RETFound-DINOv2, and RETFound-MAE. The x-axis denotes the tasks, and the evaluation metrics include AUROC. The average AUROC of DINOv2-ViT-giant, DINOv3-ViT-large, RETFound-DINOv2, and RETFound-MAE across ten tasks were 0.776, 0.778, 0.787 and 0.761 respectively. The average rankings across the ten tasks were 2.3, 2.7, 2.1, and 2.9 respectively.

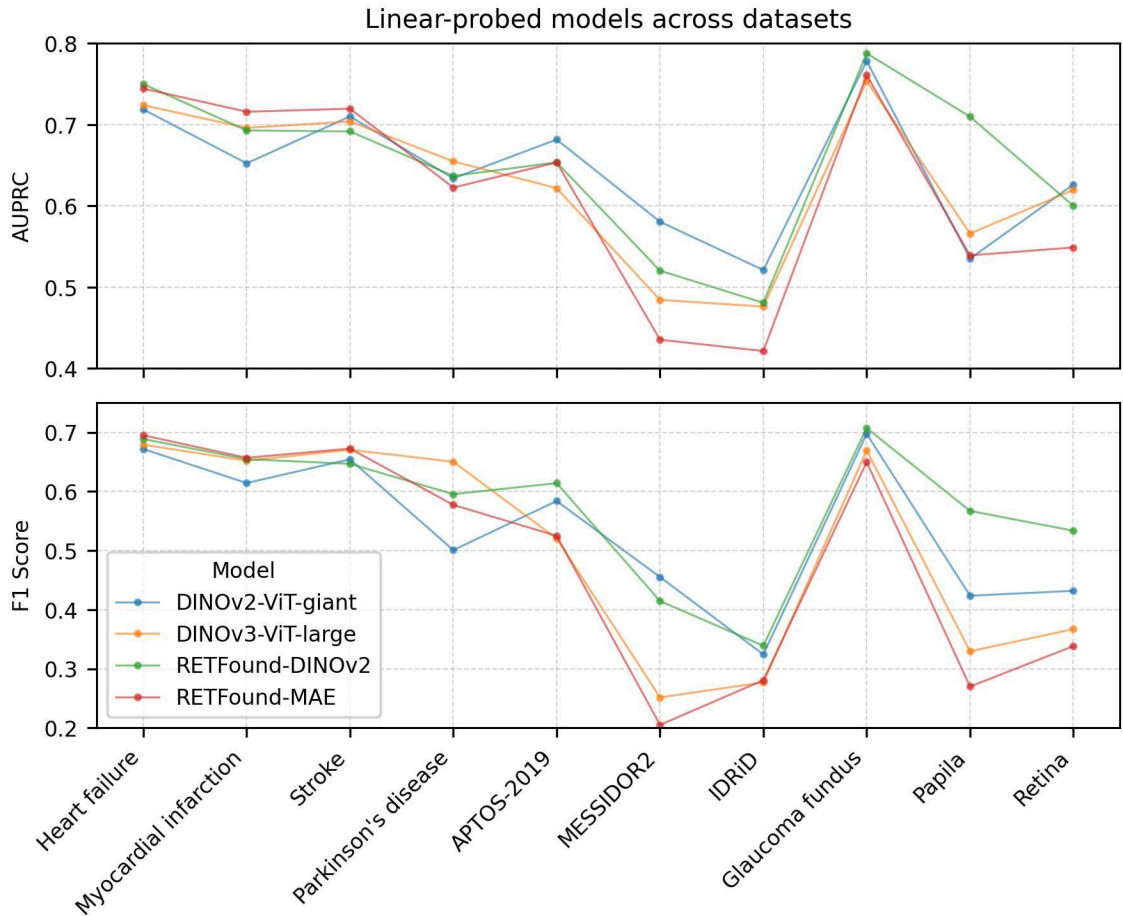

Supplementary Figure 9. Linear probing performance comparison of DINOv2-ViT-giant, DINOv3-ViT-large, RETFound-DINOv2, and RETFound-MAE. The x-axis denotes the tasks, and the evaluation metrics include AUPRC and F1 score. The average AUPRC of DINOv2-ViT-giant, DINOv3-ViT-large, RETFound-DINOv2, and RETFound-MAE across ten tasks were 0.644, 0.630, 0.652, and 0.616 respectively. The average rankings of the four models were 2.3, 2.7, 2.1, and 2.9. For the F1 score, the average values were 0.536, 0.507, 0.576, and 0.487. The average rankings of the F1 score were 2.6, 2.9, 1.7, and 2.8 respectively.

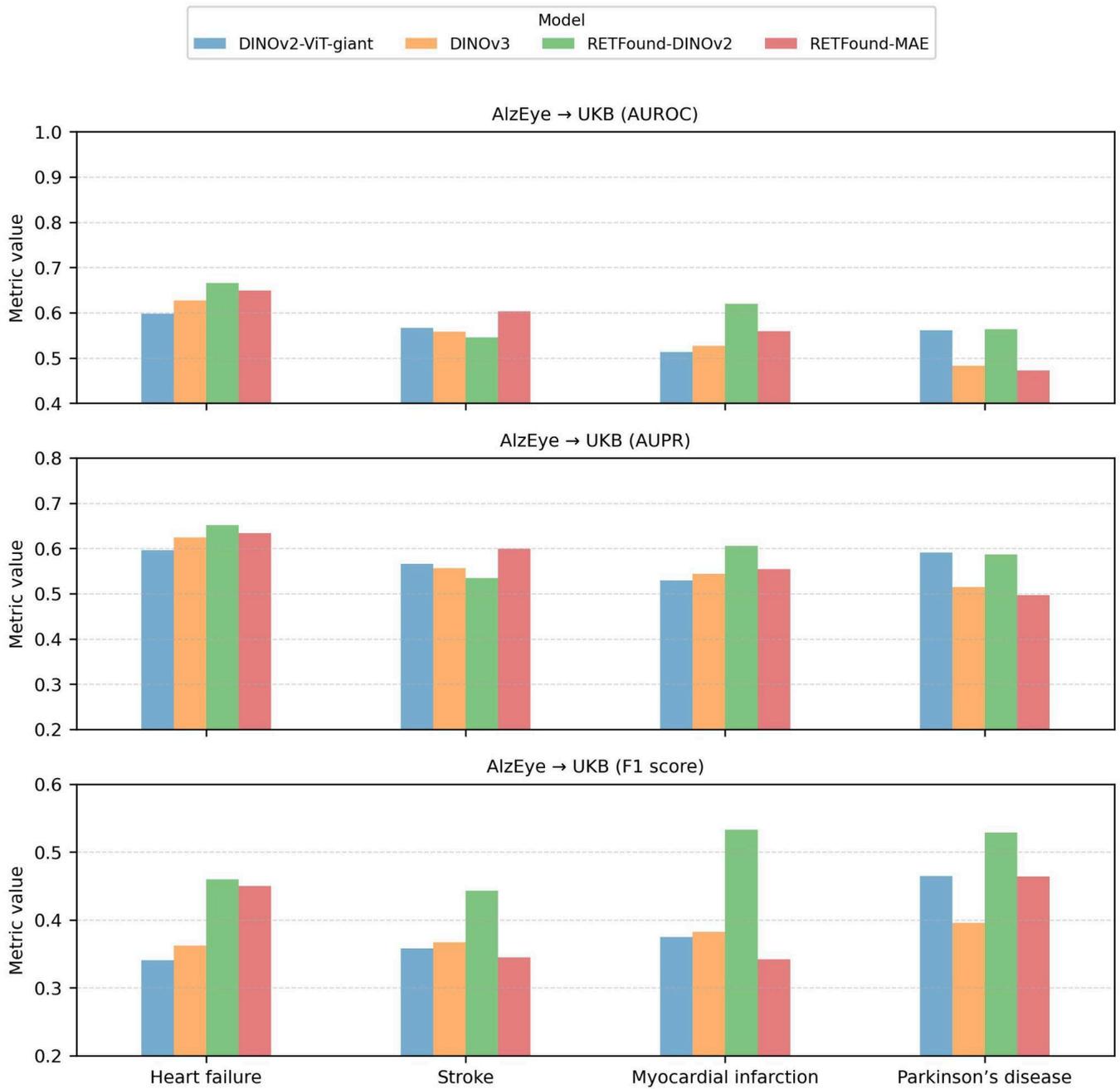

Supplementary Figure 10. Externally evaluating models fine-tuned to the AlzEye dataset on the UK Biobank dataset. The applications include heart failure, stroke, myocardial infarction, and Parkinson's disease prediction.

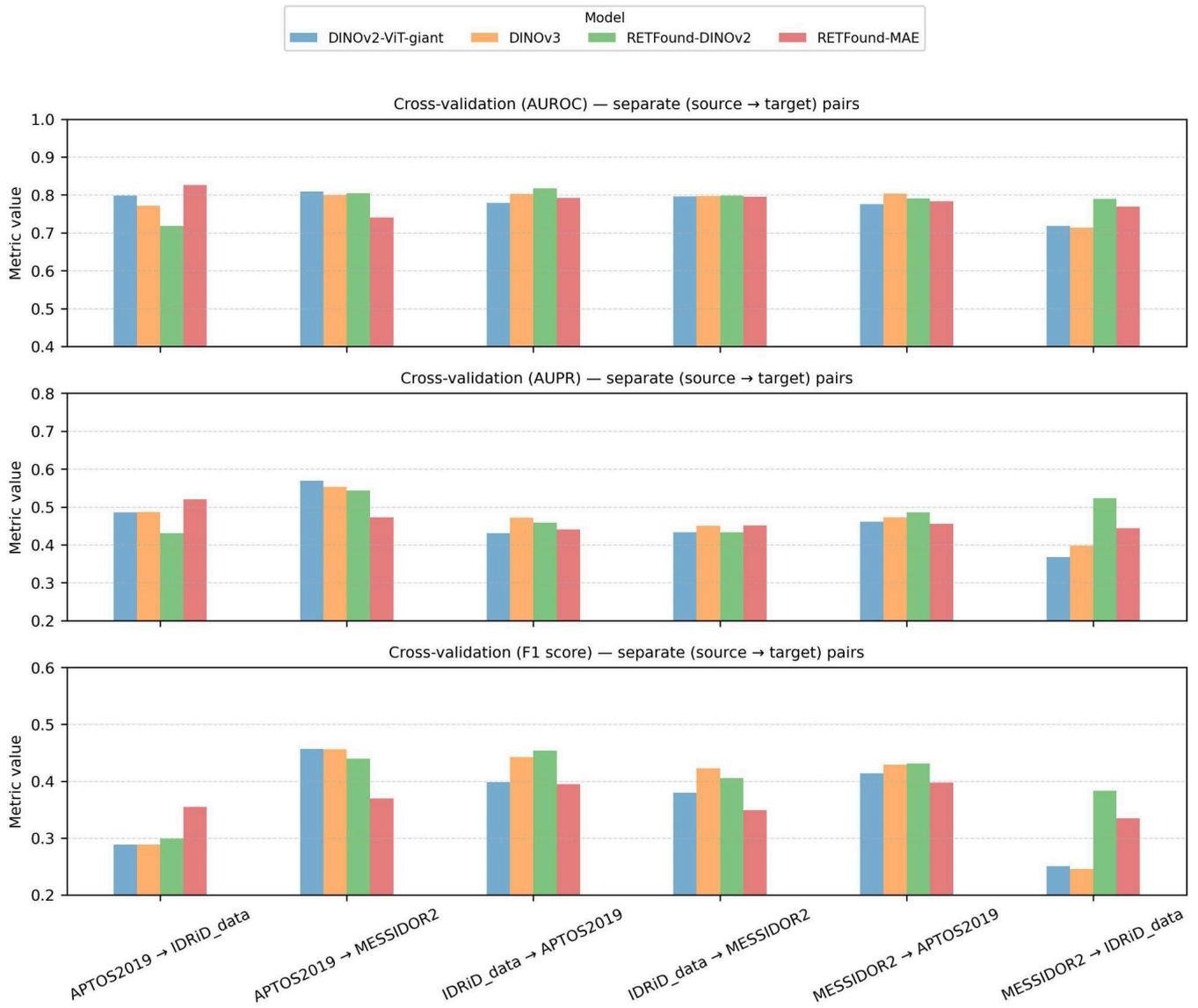

Supplementary Figure 11. Externally evaluating models via cross-validation on three diabetic retinopathy datasets. The models were fine-tuned to one dataset and externally evaluated on the other two datasets..

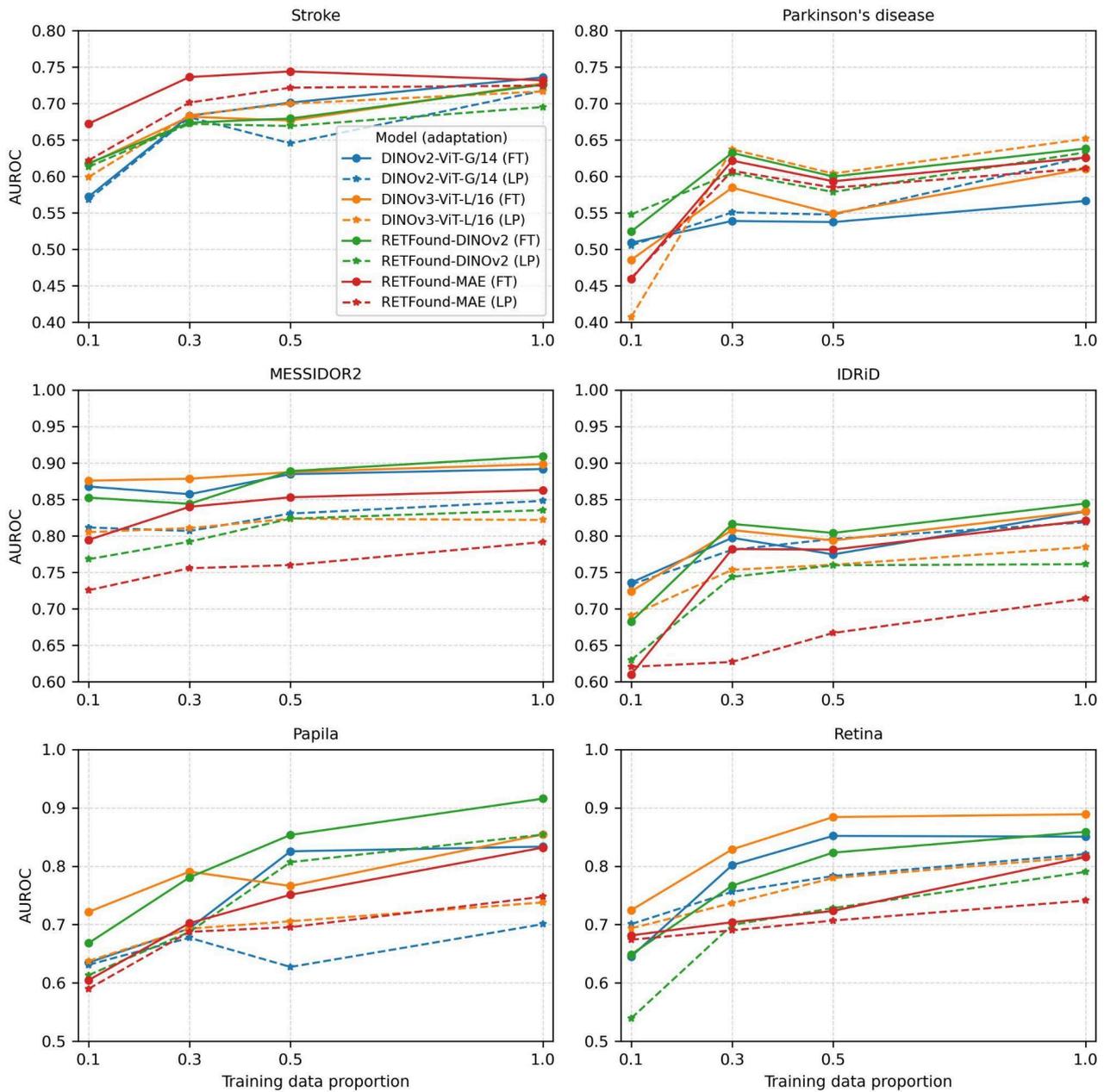

Supplementary Figure 12. Data efficiency of DINOv2-ViT-Giant, DINOv3-ViT-Large, RETFound-DINOv2, and RETFound-MAE in downstream tasks. Models were trained with varying proportions of the training data and evaluated on the full test sets. Dashed lines indicate linear probing results, while solid lines indicate fine-tuning results.

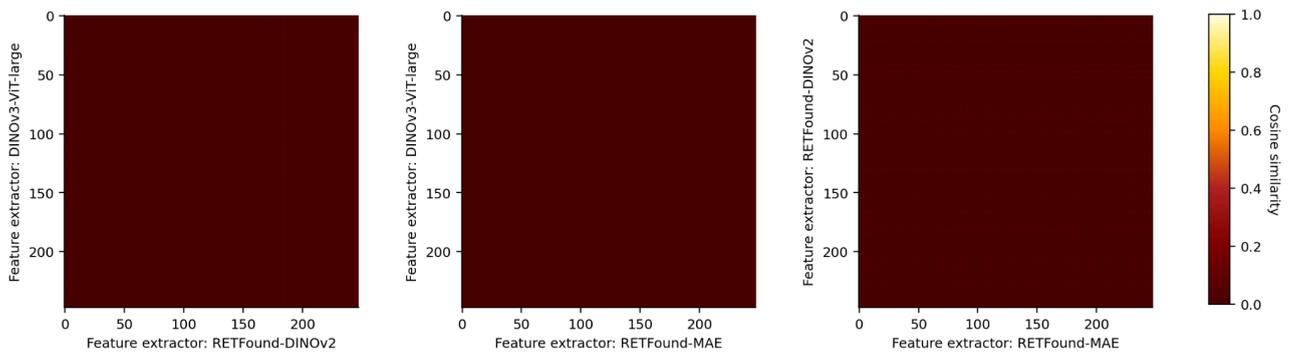

Supplementary Figure 13. Cross-model similarity heatmaps of image features extracted by DINOv3-ViT-large, RETFound-DINOv2, and RETFound-MAE. Each heatmap depicts pairwise feature similarity across retinal images, where brighter colours indicate higher similarity. All three models demonstrate low cross-model similarity across any image pairs, indicating the misalignment across their feature space. X and Y axes indicate the image ID.